\documentclass[12pt,preprint]{aastex}
\usepackage{amssymb}
\usepackage{appendix}
\usepackage{graphicx}

\shorttitle{High-redshift AGNs}

\shortauthors{Mao \& Kim}

\begin{document}


\title{On the Evolution of High-Redshift Active Galactic Nuclei}


\author{
Jirong Mao\altaffilmark{1,2,3}
and
Minsun Kim\altaffilmark{3}
}
\altaffiltext{1}{Yunnan Observatories, Chinese Academy of Sciences, 650011 Kunming,
Yunnan Province, China} 
\altaffiltext{2}{Key Laboratory for the
Structure and Evolution of Celestial Objects, Chinese Academy of
Sciences, 650011 Kunming, China}
\altaffiltext{3}{Korea Astronomy and Space Science Institute, 776,
Daedeokdae-ro, Yuseong-gu, Daejeon 305-348, Republic of Korea}

\email{jirongmao@mail.ynao.ac.cn}

\begin{abstract}
We build a simple physical model to study
the high-redshift active galactic Nucleus (AGN) evolution within the co-evolution framework
of central black holes (BHs) and their host galaxies.
The correlation between the circular velocity of a dark halo $V_c$ and the velocity dispersion of a galaxy
$\sigma$ is used to link the dark matter halo mass and BH mass. The dark matter halo mass function is converted to the BH 
mass function for any given redshift.
The high-redshift optical AGN luminosity functions (LFs) are constructed.
At $z\sim 4$, the flattening feature is not shown at the faint end of the optical AGN LF. This is consistent with observational results.
If the optical AGN LF
at $z\sim 6$ can be reproduced in the case in which central BHs have the Eddington-limited accretion, it is possible for the AGN lifetime to have a small value of $2\times 10^5$ yrs.
The X-ray AGN LFs and X-ray AGN number counts are also calculated
at $2.0<z<5.0$ and $z>3$, respectively, using the same parameters adopted in the calculation for the optical AGN LF at $z\sim 4$.
It is estimated that about 30 AGNs per $\rm{deg}^2$ at $z>6$ can be detected with a
flux limit of $3\times 10^{-17}~\rm{erg~cm^{-2}~s^{-1}}$ in the $0.5-2$
keV band.
Additionally, the cosmic reionization is also investigated.  
The ultraviolet photons emitted from the high-redshift AGNs mainly contribute to the cosmic reionization, and the central BHs of the high-redshift AGNs have a mass range of $10^6-10^8M_\odot$.
We also discuss some uncertainties in both the AGN LFs and AGN number counts originating from
the $M_{\rm{BH}}-\sigma$ relation, Eddington ratio, AGN lifetime, and X-ray attenuation in our model. 
\end{abstract}




\section{Introduction}
Active galactic nucleus (AGN) evolution is one of the vital issues to
investigate the central black hole (BH) accretion, galaxy formation, and large-scale structure of the
universe. Some statistical results, such as luminosity
functions (LFs) and number counts, can be obtained from
multi-wavelength sky surveys. The optical LFs of
quasi-stellar objects (QSOs) were obtained from the 2dF survey
\citep{boyle00,croom04}. Using the data from ROSAT, {\it{Chandra}},
and XMM-{\it{Newton}} satellites, \citet{miyaji00},
\citet{miyaji01}, and \citet{hasinger05} constructed the soft
X-ray AGN LFs in the $0.2-2$ keV band. The hard X-ray LFs and hard
X-ray cosmic background in the $2-10$ keV band were studied as
well \citep{cowie03,ueda03,barger05,lafranca05}. From these early
studies, we know that neither density evolution nor luminosity
evolution can fully describe AGN LF evolutionary properties.
However, a downsizing feature has been reported, which is the number density of galaxies/AGNs
peaking at low redshift \citep{cowie96,lemaux14,ueda14}. Thus, the crucial explorations of the AGN LFs in the optical band for the high-redshift universe,  which have been performed during recent years, have become more important. The QSO LFs at $3.6<z<5.0$
from the Sloan Digital Sky Survey (SDSS) have been given \citep{fan01}.
The AGN LFs at $3.5<z<5.2$
have been extended to the faint end of $M_{1450}\sim -21.5$ by
the great observatories origins deeps survey (GOODS)
high-redshift AGN sample \citep{cristiani04,fontanot07}.
The faint type-1 AGN
LFs have been given by \citet{bongiorno07} from the VIMOS-VLT deep survey.
With the 2dF-SDSS data, the AGN LF at $0.4<z<2.6$
has been extended 2.5 mag fainter \citep{croom09} than the previous
one. At $z\sim 4$, some faint end data points for QSO LF have been
provided by the NOAO deep wide-field survey and the deep lens
survey \citep{glikman10}. The AGN LFs at $z\sim 6$ were obtained by the SDSS deep
survey \citep{jiang09} and the Canada-France high-redshift QSO
survey \citep{willott10}. The Subaru high-redshift QSO survey
has also discovered some faint AGNs at $z\sim 6$ \citep{kashikawa14}.
High-redshift AGN detections in the X-ray band are also exciting. From the {\it XMM-Newton}
medium-sensitivity survey, \citet{ebrero09} have derived AGN LFs
($0<z<3$) which can be described by a luminosity-dependent
density evolution (LDDE) model. The AGN LF at $z\leq 5$ obtained from a faint
{\it Chandra} sample has been reported by \citet{yencho09}.
\citet{aird10} have found the high space density of low-luminosity
AGNs at $z>2$ from the 2 Ms {\it Chandra} deep survey and the AEGIS-X 200
ks survey. The number counts of X-ray selected QSOs at $z>3$
have been obtained from the COSMOS survey and the Subaru/{\it XMM-Newton} deep survey
\citep{brusa09,civano11,hiroi12}.
The number density evolution of obscured AGNs has been
mentioned by \citet{fiore08,fiore09}, \citet{brusa10}, and \citet{hiroi12}. AGNs
with column densities above $10^{24}~\rm{cm^{-2}}$ are denoted as
``Compton thick'', and these heavily obscured AGNs represent a key phase
of AGN-host galaxy co-evolution. It is quite interesting that
two Compton-thick AGNs have been
discovered in the {\it XMM}-{\it Newton} deep survey ($z=3.7$) and the 4 Ms {\it Chandra}
deep field (CDF) survey ($z=4.76$), respectively \citep{comastri10a,gilli11}. The Compton-thick AGN surveys
have been done by \citet{brightman14} and \citet{lanzuisi14}. The AGN number density and the cosmic X-ray background
have been studied with large observational samples \citep{ueda14,aird15,miyaji15}.

Some problems with the AGN LF have been raised by these new observations. For example, at $z>4$,
a flattening feature at $M_{1450}>-23$ seen in the optical AGN LF
\citep{fontanot07} was explained by the physics of central
BH growth and AGN/supernova feedback \citep{lapi06,fontanot06}.
However, the number density of faint optical AGNs ($-23<M_{1450}<-21$) reported by
\citet{glikman10}  is about 10 times larger
than that of \citet{fontanot07}. With additional spectral identifications,
\citet{glikman11} re-built the AGN LF at $z=4$. However, 
the number density of these faint AGNs is still two times
that given by \citet{fontanot07}. This observational discrepancy was also mentioned by \citet{fiore10}.
One possible reason for this
discrepancy is as follows \citep{glikman10,glikman11}. \citet{fontanot07} used
the QSO spectral template of \citet{cristiani90}, which empirically determines
a distribution of continua blueward of Ly$\alpha$, and the power-law spectral template
has a slope of 0.7; \citet{glikman10,glikman11} had the QSO spectral library
template modified by a composite using the Gaussian distribution of power-law continua, and a slope
of 0.5 is given in the spectral template. 
If the sample selected by \citet{glikman11} has better completeness than that selected by \citet{fontanot07},
the flattening feature given by \citet{fontanot07} would be significantly reduced.
While Masters et al. (2012) built the optical LFs at $3<z<5$ from the COSMOS survey field. The space density of faint AGNs at $z\sim 4$ is lower than the result of Glikman et al. (2011) by a factor of $3\sim 4$. They suggested that this discrepancy may be due to the contamination of dwarf stars and high-z galaxies in the sample of Glikman et al. (2011), and spectroscopic follow-up identification is still challenging for this contamination constraint.
Therefore, the final high-redshift AGN candidates
are different due to different selection functions.
Although the AGN LF at $z\sim 6$ has been extended to the faint end of $M_{\rm{1450}}\sim -22$, it seems that the AGN LF is still
steep at the faint end \citep{willott10, kashikawa14}.

One widely accepted issue of AGN evolution is that the AGN and its host galaxy have
co-evolution characteristics \citep{richstone98}. A strong correlation was found between the central BH
mass $M_{\rm{BH}}$ of a normal galaxy and the velocity dispersion
$\sigma$ of a galactic bulge \citep{ferrarese00,gebhardt00,tremaine02,gultekin09,batcheldor10}.
This $M_{\rm{BH}}-\sigma$ relation (presented as
$logM_{\rm{BH}}=\alpha+\beta log(\sigma/200~\rm{km~s^{-1}})$, where $\alpha$
and $\beta$ are fitting parameters) is
also valid for the AGN population
\citep{shields03,shields06,greene07,wu07,woo10}. A correlation
between a BH mass $M_{\rm{BH}}$ and the galactic spheroid mass $M_{\rm{sph}}$
\citep{magorrian98,marconi03,haring04,nelson04} also indicates the
co-evolution feature of AGNs and their host galaxies. Moreover,
the $M_{\rm{BH}}-\sigma$ relation evolves with redshift, and
the ratio between the $M_{\rm{BH}}$ and $M_{\rm{sph}}$ increases with redshift
\citep{shields06,mclure06,decarli10}. These evolution properties
provide the possibility that the central BH growth could finish
before its host spheroid formation. In other words,
massive BHs grow rapidly in the early universe without
commensurate growth of their host galaxies \citep{grupe04,shields06}. 
On the contrary, Schulze \& Wisotzki (2014) performed a fitting procedure of the BH-bulge relation with the Monte Carlo simulation test, in which the observational selection effects were considered. They did not find any cosmological evolution of the BH-bulge relation when the selection effects were corrected.
On the other hand,
\citet{loeb03} suggested that the circular velocity $V_{\rm{c}}$ of a dark matter halo potential
well can be linked to the velocity dispersion $\sigma$ of a galactic
bulge, and the theoretical relationship between
$\sigma$ and $V_{\rm{c}}$ was also examined by
observational data \citep{ferrarese02}. Therefore, we can link
the physical processes of dark matter halos to those of baryons
at a certain redshift.

Theoretical investigations of co-evolution have been carried out between central BHs and their host galaxies \citep{silk98,fabian99,king03,king05,begelman05,croton09,nayakshin09,king10}.
\citet{shankar09a} predicted the velocity
dispersion functions of spheroid galaxies at
$0<z<6$. Their results indicate a redshift evolution of the $M_{\rm{BH}}-\sigma$
relation. Similar evidence for the evolution of the $M_{\rm{BH}}-\sigma$
relation was given by numerical simulations \citep{robertson06}.
The redshift evolution of a $M_{\rm{BH}}/M_{\rm{sph}}$ was also
modeled \citep{somerville09,lamastra10}. \citet{haehnelt93}
presented a QSO evolution model in which massive BHs are formed
inside virialized dark matter halos. AGN LFs were
studied within the framework of co-evolution between central BHs
and their host galaxies, and the BH accretion history was explored in
detail \citep{haehnelt98}. A linear relation between a dark matter
halo mass and BH mass was used to derive the AGN LF
\citep{haiman98,wyithe02}, and the processes of BH self-regulated
growth were also considered \citep{wyithe03}. Several complicated
models of QSO activity, including the BH accretion process
\citep{mahmood05} and the BH merger process \citep{shen09}, were given as
well. From numerical simulations, \citet{hopkins05} proposed a
luminosity-dependent AGN lifetime to reproduce the AGN LFs.
The AGN/supernova feedback processes were used to explain the
observational downsizing feature of the AGNs/BHs
\citep{croton06,fontanot06,lapi06,lapi14,marulli08,menci08}.
\citet{lapi06} noticed that AGN shining occurs
after dark matter halo virialization due to the growth of
a central BH seed. Thus, a time delay between dark matter halo
virialization and AGN shining should be considered. In this case,
super-Eddington accretion is important for the central BH growth.

Based on a semi-analytic model of the co-evolution between BHs and their hosts, Fanidakis et al. (2012) investigated the AGN evolution, considering both the thin-disk accretion and the ADAF model.
The optical and X-ray AGN LFs were reproduced when the AGN obscuration form of Hasinger (2008) was taken into account. This model predicted a hierarchical buildup of BH mass, but the downsizing feature for different AGN populations was clearly shown due to the different accretion modes and the AGN obscuration. Hirschmann et al. (2012) comprehensively investigated some accretion modes. Standard accretion, varing sub-Eddington limit for the maximum accretion, and disk instability accretion were considered. The BH seed mass for the merger process was also considered. A cosmological numerical simulation for the BH growth was performed by Hirschmann et al. (2014). The downsizing feature of the AGN number density evolution was well explained when the gas density nearby the massive BH was included in their model. Thus, the processes of star formation and AGN/supernova feedback should be also involved. The star formation process was further explored by Enoki et al. (2014). In their semi-analytic model, the starburst is caused by an AGN merger, and the cold gas accretion triggers AGN shining. The amount of cold gas for the BH accretion decreases with cosmic time, and the AGN lifetime decreases with redshift. Then, the downsizing trend of the AGN evolution is shown. From above theoretical models, the observed downsizing feature of the AGN evolution can also be explained under the framework of the hierarchical structure formation.

Some research efforts have been attempted toward explaining 
the faint end of the high-redshift AGN LFs. \citet{hopkins07} determined
the bolometric AGN LF at $0<z<6$ using
a large set of AGN LFs in different rest-frame wavelengths. The
faint-end slope of the AGN LF was confirmed. \citet{shankar09,shankar10,shankar13}
comprehensively studied bolometric LFs through the arrangements of BH
accretion, radiative efficiency of AGNs, and AGN clustering. However,
it was pointed out by \citet{willott10} that those LFs built by
\citet{hopkins07} and \citet{shankar09} overestimated the number
density of faint AGNs at $z=6$. For instance, in a luminosity range
of $-25.5<M_{1450}<-24.5$, the AGN number density calculated by \citet{shankar09} is
about five times larger than those given by \citet{jiang09}
and \citet{willott10b}. Moreover, the predictions of high-redshift
AGN number counts in the X-ray band are also uncertain.
\citet{rhook08} proposed a dark matter halo merger model
to drive AGN shining.
\citet{marulli08} constructed a semi-analytic model of
the co-evolution between BHs and host galaxies. It was assumed that central BHs
continuously accrete surrounding hot gas for AGN shining. As reported by \citet{gilli10}, at $z>6$,
with a flux limit of $3\times 10^{-17}~\rm{erg~cm^{-2}~s^{-1}}$ in the
$0.2-2$ keV band, an optimistic detection of $500-600$
AGNs per $\rm{deg}^2$ and a pessimistic detection of $\sim$ 16
AGNs per $\rm{deg}^2$ were predicted by \citet{rhook08} and
\citet{marulli08}, respectively.
Because many obscured low-luminosity AGNs can be found at high redshifts
\citep{lafranca05,hasinger08, menci08}, 
high-redshift
AGN detections in the X-ray band are strongly encouraged by some
satellite proposals. For example, the Wide Field X-ray Telescope
(WFXT) can perform deep
observations in 100 $\rm{deg}^2$ to a flux limit of $3\times
10^{-17}~\rm{erg~cm^{-2}~s^{-1}}$ in the $0.5-2$ keV band \citep{rosati10}. Those
unobscured AGNs beyond redshift 6 can be effectively detected.  Some obscured AGNs can also be explored up to
$z\sim 4$ \citep{comastri10a,gilli10}. Therefore, modeling
predictions of the high-redshift AGN number density in the X-ray band
could also be important for future missions.

The ultravoilet photons emitted by AGNs can effectively ionize the
intergalactic medium (IGM).
High-redshift AGN LFs are used to study cosmic reionization.
The so-called Gunn-Peterson trough \citep{gunn65} seen in the optical
spectra of the SDSS AGN sample provides excellent evidences to constrain the cosmic
reionization epoch \citep{fan02,fan06}.
It was estimated by
\citet{fan06b} that the QSO contribution to the ionizing
background is less than 30\% of the star-forming galaxy contribution.
The AGN contribution to the cosmic reionization \citep{shankar07}
was revisited by \citet{glikman10}. However, the Subaru high-redshift quasar survey has shown that the AGNs at redshift 6
still do not have enough photons to ionize the universe \citep{kashikawa14}. In this paper, as a consequent issue of the  high-redshift AGN evolution,
some rough estimations for the cosmic reionization are given.

We build a simple analytic model to study high-redshift AGN evolution within the framework of joint evolution between
central BHs and their host galaxies. Beginning from the dark matter halo
mass function, we derive the AGN LFs by applying the 
$M_{\rm{BH}}-\sigma$ and $V_c-\sigma$ relations.
High-redshift AGN number counts in the X-ray band are obtained as
well. We compare our results to those observational data.
The cosmic reionization is also examined.
In our physical recipe, the Eddington
ratio $\lambda$, AGN lifetime $\tau_{\rm{AGN}}$, $\alpha$, and
$\beta$ in the $M_{\rm{BH}}-\sigma$ relation, are free parameters
(see Table 1 for details). All of these parameters
are within their observational constraints.
We do not perform optimized 
fittings to those observational AGN LF data by adjusting the parameters in 
our model. Instead, we choose reasonable values for
these parameters to produce the AGN LFs and AGN number counts by comparison with the observational data. Thus, 
some related physical processes can be further explored.
Section 2.1 describes our physical model. The AGN LFs and AGN number counts
are derived in Section 2.2. In Section 3, the contributions from
the high-redshift AGNs to the cosmic reionization are estimated.
Discussion and conclusions are given in Sections 4 and 5, respectively.
Throughout the paper, we adopt the following cosmological parameters:
$H_0=72~\rm{km~s^{-1}~Mpc^{-1}}$, $\Omega_{\rm{m}}=0.3$, and
$\Omega_\Lambda=0.7$.

\section{High-Redshift AGN Evolution}
\subsection{Model Description}
AGN LF is one of the powerful tools to investigate central
BH accretion and host galaxy evolution. Some examples of the AGN LF
formulism have been derived
\citep{wyithe02,wyithe03,lapi06,shen09}. In this paper, we derive
the AGN LF as follows:
\begin{equation}
\Phi(L,z)=\frac{dn_{\rm{AGN}}}{dlogL}=\frac{d^2N}{dM_{\rm{halo}}dz}\frac{dM_{\rm{halo}}}{dM_{\rm{BH}}}\frac{dM_{\rm{BH}}}{dLogL}\frac{dz}{dt}\tau_{\rm{AGN}},
\end{equation}
where $M_{\rm{halo}}$ is the mass of a dark matter halo,
$M_{\rm{BH}}$ is the mass of a BH, $L$ is the luminosity in a
certain observational band, $dN/dM_{\rm{halo}}$ is the
dark matter halo mass function.
We further calculate the time derivative of $dN/dM_{\rm{halo}}$ (Sasaki 1994; Kitayama \& Suto 1996). The result contains two terms, which are formation and destruction terms. Here, we only keep the formation term written as $d^2N/dM_{\rm{halo}}dz$ and ignore the destruction term. Therefore, it is assumed that an AGN is formed inside a newly formed dark matter halo at a certain redshift, and the dark matter halo merger effect is not included in our model. We also assume that dark matter halos are virialized to harbor AGNs. 
The relation of $M_{\rm{halo}}$ and $M_{\rm{BH}}$ is given by
$dM_{\rm{halo}}/dM_{\rm{BH}}=(dM_{\rm{halo}}/dV_c)(dV_c/d\sigma)(d\sigma/dM_{\rm{BH}})$,
where $V_c$ is the circular velocity of the dark matter halo potential well, and
$\sigma$ is the velocity dispersion of the host galaxy.
The AGN lifetime $\tau_{\rm{AGN}}$ is far less than the cosmic time. Therefore, 
to derive AGN LF, seven physical issues are involved in our
procedure: the dark matter halo formation rate and the $M_{\rm{halo}}-V_c$
relation are related to the dark matter halo mass function and
the dark matter halo potential well, respectively; the $V_c-\sigma$
relation is applied to link the dark matter halo mass to the BH
mass; the $M_{\rm{BH}}-\sigma$ relation indicates the relation between
the central BH and the host galaxy; the BH accretion is related to the physical
properties of the central BH; the bolometric correction should be
considered when we transfer bolometric luminosity to a specific 
luminosity in the optical/X-ray band; and the AGN lifetime is the AGN shining
timescale in either the optical or X-ray band. In the following
subsections, we describe these physical points adopted in our model
in detail. 

\subsubsection{Link between Dark Matter Halos and BHs}
The dark matter halo mass function
$dN/dM_{\rm{halo}}$ at a certain redshift can be
described by the Press-Schechter function \citep{press74}. Here, we
use the $dN/dM_{\rm{halo}}$ form given by
\citet{sheth99,sheth02}, which is more accurate than the 
Press-Schechter function at high-redshift \citep{springel05}.
The dark matter halo formation rate
$d^2N/dM_{\rm{halo}}dz$ can also be calculated \citep{sasaki94,kitayama96}.
The circular velocity of the dark matter halo can be obtained as
$V_{c}=
375(M_{\rm{halo}}/10^{12}~M_\odot)^{1/3}((1+z)/7)^{1/2}~\rm{km~s^{-1}}$
\citep{loeb03}. With a certain density profile of the dark matter
halo, \citet{loeb03} extrapolated $V_c$ from the virial radius of
a dark matter halo to the effective radius of a galaxy, and a factor
of $1/\sqrt{2}$ was multiplied to this circular velocity at the
effective radius. The final result was regarded as the
line-of-sight velocity dispersion of the galaxy. We use their results
and choose the Navarro-Frenk-White (NFW) density profile of the dark
matter halo \citep{navarro97}. Thus, an approximation relation
$\sigma=0.69V_c$ is obtained. This theoretical result is
consistent with the linear fit of $\sigma=(0.65\pm0.02) V_c$ from the
observational data \citep{ferrarese02}. Applying this
theoretical $V_c-\sigma$ relation and the $M_{\rm{BH}}-\sigma$
relation described in the next subsection, we can link the dark matter
halo mass to the BH mass. 

We use the $V_c-\sigma$ relation to link the BH mass
and dark matter halo mass. However, some data points from the 
$\sigma$ and $V_c$ measurements of bulgeless galaxies
\citep{kormendy11} have a strong deviation to the previous
$V_c-\sigma$ relation given by \citet{ferrarese02}. Here, we
mention that the BH mass considered in this paper is larger than
$10^6$ $M_\odot$. Most bulgeless galaxies selected in the work of
\citet{kormendy11} have a velocity dispersion less than
$60~\rm{km~s^{-1}}$. This velocity dispersion put a constraint on the central BH mass, which is less
than $10^6~M_{\odot}$. Therefore, these low-mass central BHs and
related low-mass galaxies are not within our consideration of AGN
LFs and AGN number counts. Moreover, \citet{volonteri09}
found that low-mass bulgeless galaxies are unlikely to host
central BHs. Furthermore, \citet{volonteri11} re-examined
those observational data. They reported that the $M_{\rm{BH}}-V_c$
correlation may be as strong as the $M_{\rm{BH}}-\sigma$ relation.
Thus, we think that the correlation between the
central BH and dark matter halo is valid to derive the AGN LFs and AGN
number counts at high redshifts.
In the following subsections,
we focus on the physical processes of both central BHs and their host galaxies, which are related to AGN LFs.

\subsubsection{Co-evolution between BHs and Their Host Galaxies}

Central BH accretion and AGN activity are strongly related to
the AGN environment and host galaxy evolution
\citep{silk98,fabian99}. Our AGN LF model is within the
framework of co-evolution between AGNs and their host galaxies.
Some important issues in our model are the
$M_{\rm{BH}}-\sigma$ relation, BH accretion, bolometric
correction, and AGN lifetime. To
derive the AGN LFs, the $M_{\rm{BH}}-\sigma$ relation is required. This
relation with a form of $M_{\rm{BH}}\propto \sigma^5$ was first
proposed by \citet{silk98}. It indicates a strong co-evolution
between central BHs and their host galaxies. Because the AGN feedback has been
considered, the form of $M_{\rm{BH}}\propto \sigma^4$ can be
derived \citep{king03,king05}. In principle, because the
$M_{\rm{BH}}-\sigma$ relation originates from a dynamic process,
it should be independent of the observational wavelength.
\citet{robertson06} examined the $M_{\rm{BH}}-\sigma$ relation with the numerical simulation
model. In their model, the circular velocity $V_c$ is derived from the NFW dark matter halo profile.
A certain gas fraction is given in each halo, and the BH seed grows through Bondi accretion.
Especially, the feedback from both supernova and accreting BH is included.
The final $M_{\rm{BH}}-\sigma$ relation can be derived from a large set of hydrodynamic simulations
with additional physics, such as the equation of state of the interstellar medium, galactic surface mass density, and dark matter concentration. Therefore,
the $M_{\rm{BH}}-\sigma$ relation of
$logM_{\rm{BH}}=8.01+4.02log(\sigma/200~\rm{km~s^{-1}})-0.138log(1+z)$
computed by \citet{robertson06} is used in our model to
reproduce the AGN LFs in the optical and X-ray bands. It is
worth noting that this relation is dependent on the redshift. As
mentioned in Section 1, 
the modeling results and the observational measurements for the $M_{\rm{BH}}-\sigma$ relation are roughly consistent.
However, the observational $M_{\rm{BH}}-\sigma$ relation has an
intrinsic scatter of $0.25-0.3$ dex \citep{tremaine02}.
\citet{gultekin09} investigated the intrinsic scatter of
the $M_{\rm{BH}}-\sigma$ relation in detail and re-examined the
$M_{\rm{BH}}-\sigma$ relation using updated data. As mentioned in
their paper, there are systematic uncertainties when measuring the BH
mass in a host galaxy. In this paper, we consider the effect of this intrinsic
scatter in the $M_{\rm{BH}}-\sigma$ relation. The best-fit of the $M_{\rm{BH}}-\sigma$ relation from
\citet{gultekin09} is $log(M_{\rm{BH}}/M_{\odot})=(8.12\pm
0.08)+(4.24\pm 0.41)log(\sigma/200~\rm{km~s^{-1}})$. The intrinsic
root-mean-square scatter of $log(M_{\rm{BH}}/M_{\odot})$ is given
as $0.44\pm 0.06$, and this intrinsic scatter is assumed to be
independent of $\sigma$. Taking into account the effect from this
intrinsic scatter of the observational $M_{\rm{BH}}-\sigma$ relation,
we can reproduce the AGN LFs again. Thus, two kinds of AGN LFs, one
derived by adopting a theoretical $M_{\rm{BH}}-\sigma$ relation
and the other by adopting an observational
$M_{\rm{BH}}-\sigma$ relation with an intrinsic scatter, are
compared.

The accretion property of the central BH is a key point to produce the AGN
bolometric luminosity. The Eddington ratio of accretion is defined
by $\lambda=L_{\rm{bol}}/L_{\rm{edd}}$, where $L_{\rm{bol}}$ is
the bolometric luminosity, and $L_{\rm{edd}}$ is the Eddington luminosity.
The value of $\lambda$
has a wide distribution with its peak range from 0.01 to 1
\citep{hopkins09a,shen09,fanidakis10}. In the sub-Eddington
accretion case of $\lambda \sim 0.5$,
the bolometric AGN LF modeled by
\citet{shankar09} is about three to five times larger than
observational ones at $z=6$ within a luminosity range of $-25.5<M_{1450}<-24.5$
\citep{willott10}. While the Eddington ratio has been constrained as $\lambda\sim 1$ at $z=6$ by the
observations of \citet{willott10} and \citet{kashikawa14}.
To investigate this problem, we try two accretion modes.
We first adopt a sub-Eddington accretion of $\lambda=
0.25(M_{\rm{BH}}/2\times 10^9M_\odot)^{0.5}$ to reproduce the AGN LFs and AGN number counts in general.
We also adopt a
mode of $\lambda=2.0(M_{\rm{BH}}/2\times 10^9M_\odot)^{0.5}$ to particularly reproduce the optical AGN LFs at $z\ge 6$.
The accretion mode as a function of BH mass was proposed by Shankar et al. (2009b). Although the low-mass BH growth at low redshift is suppressed with this accretion mode, the growth of high-mass BHs at high redshift can be well estimated. Shankar et al. (2009b) got somewhat better fits to the BH mass function in general with this accretion mode. From the observational point of view, most detected high-redshift AGNs have massive BHs with high accretion rates. In order to reproduce high-redshift AGN LFs, we focus on high-mass BHs with high accretion ratios. Thus, this accretion mode is adopted in our task.
In Section 4, we further provide observational constraints and
physical explanations on the accretion modes adopted in this work.

We perform the bolometric corrections, $k_B=L_{bol}/L_B$ and
$k_x=L_{bol}/L_x$, to convert the bolometric luminosity $L_{bol}$ to
the B-band luminosity $L_B$ and the X-ray luminosity $L_x$, respectively.
\citet{elvis94} determined two observational values, $k_B=11.4$ and
$k_x=35$, for broad-line AGNs. $k_x=85$ was determined in the case in which AGNs have no broad-line feature\citep{barger05}. Here, we take the bolometric
correction results of \citet{marconi04}: $log[L/L_x(2-10
~{\rm{keV}})]=1.54+0.24(logL-12)+0.012(logL-12)^2-0.0015(logL-12)^3$,
$log[L/L_x(0.5-2
~{\rm{keV}})]=1.65+0.22(logL-12)+0.012(logL-12)^2-0.0015(logL-12)^3$,
and
$log(L/L_B)=0.80-0.067(logL-12)+0.017(logL-12)^2-0.0023(logL-12)^3$,
where $L$ is the bolometric luminosity in units of $L_\odot$. The
template spectrum of \citet{marconi04} constructed by the
intrinsic AGN spectral energy distribution (SED) enables us to
precisely determine the AGN luminosities in different energy bands.

The AGN e-folding time given by the Salpeter timescale
\citep{salpeter64,marconi04} is usually considered as the AGN lifetime. From
AGN clustering measurements, we can obtain the AGN lifetime with a
wide range from $10^6$ to $10^8$ years \citep{haiman01,martini01}.
\citet{haiman02} took the value of
$\tau_{\rm{AGN}}\sim 10^7$ yrs as the AGN lifetime for high-redshift QSOs. In the
AGN-host galaxy joint evolution framework, we suggest different
AGN lifetimes for different observational energy bands. At the initial
coevolutionary stage of the central BHs and their host galaxies, AGN X-ray
emission can be detected, but AGN optical emission is still fully
absorbed. At a later stage, the AGN can be observed in the optical band when the AGN
surrounding medium is transparent. Thus, the lifetime of the X-ray
selected AGNs could be longer than that of the optically selected
AGNs. Here, we take $2\times 10^7$ and $9\times 10^7$ years
as reference values for the AGN lifetime to calculate
the optical and X-ray AGN LFs, respectively. We take an exceptional case, in which the AGN lifetime is given as $2\times 10^5$ yrs, to reproduce the optical AGN LFs at $z\ge 6$.

\subsection{LFs and Number Counts}
Assuming a sub-Eddington accretion of $\lambda= 0.25(M_{\rm{BH}}/2\times
10^9M_\odot)^{0.5}$, we use the $M_{\rm{halo}}-V_c$, $V_c-\sigma$, and
$M_{\rm{BH}}-\sigma$ relations to derive the AGN LFs. As mentioned in
Section 2.1, these relations adopted in our calculations are
obtained from theoretical works. The AGN LFs derived
from an observational $M_{\rm{BH}}-\sigma$
relation are also given in this paper, and those effects from
the intrinsic scatter of the observational $M_{\rm{BH}}-\sigma$
relation are investigated. Below, we present our results in
detail.

\subsubsection{AGN LFs in the Optical and X-Ray Bands}

We show the optical AGN LFs at $z=4$ in Figure 1. The deep
GOODS observation with a sky area of 320 $\rm{arcmin}^2$
provides a faint ($M_{1450}\le -21$) AGN sample
\citep{fontanot07}. The AGN LF from the NOAO Deep Wide-Field
Survey and the Deep Lens Survey with a total area of 3.76
${\rm{deg}}^2$ \citep{glikman10} is also shown. We clearly see
that the AGN LF from these GOODS data has lower values than the AGN LF from
\citet{glikman10} at $M_{1450}>-23$. With a sub-Eddington accretion of
$\lambda= 0.25(M_{\rm{BH}}/2\times 10^9M_\odot)^{0.5}$ and an AGN
lifetime of $2\times 10^7$ yrs, we calculate the optical AGN LF at
$z=4$, and our result is consistent with that of \citet{glikman10}.
As we described in Section 1, the
discrepancies shown among the LFs of \citet{glikman10,glikman11}, \citet{fontanot07}, 
and \citet{masters12} could be due to the different selection functions used to select high-redshift
AGN candidates. The spectral identification is challenging to identify
faint AGNs at high redshift. Moreover, at present, large error bars are given at the
faint end of both LFs from \citet{glikman10} and \citet{fontanot07}.
Thus, the discrepancy has large uncertainty.

We show the optical AGN LFs at $z=6$ in Figure 2. Six QSOs down to
a magnitude of $M_{1450}\sim -25$ were found from the SDSS deep stripe
\citep{jiang09}. The result from the Canada-France high-z quasar
survey \citep{willott10b} is also shown. The data point
at the faintest magnitude of $M_{1450}\sim -22.2$ was estimated by the
single quasar from the Subaru/XMM-{\it Newton} Deep Survey. Data points from the Subaru high-redshift quasar survey
\citep{kashikawa14} are also shown.
With the same physical parameters used to reproduce the AGN LF at $z=4$, we compute
the AGN LFs at $z=6$, and the results are shown in Figure 2. In addition, our predictions for the 
optical AGN LFs at $z=8$ and $z=9.5$ are shown in Figure 3.

The AGN LFs estimated by our model are roughly consistent with those
from updated observational data at $z=4$.
At $z=6$, our results seem to be consistent with
the observational data given by \citet{jiang09} in a luminosity range of
$-28<M_{1450}<-24$. In general, we do not have the overestimation problems that were mentioned
by \citet{hopkins07} and \citet{shankar09}.
However, our AGN LF at the faint end of $M_{1450}=-22.2$ is 10
times larger than the observational one. We note that only a
single quasar was used to build that data point \citep{willott10,willott10b}. Therefore, at
$z=6$, the observational AGN LF at the faint end is still poorly
constrained, although the observational AGN LF still seems steep at the faint end when we consider the data points of the Subaru
high-redshift quasar survey \citep{kashikawa14}. From our results, we expect that more faint
AGNs can be detected by some wide-field surveys of the Subaru telescope, such as the Hyper Suprime-Cam (HSC)
survey and the Prime Focus Spectrograph (PFS) survey, in the future.

The AGNs at $z=6$ have been identified by \citet{willott10,willott10b} and \citet{kashikawa14},
 and the central BHs with a mass of $10^8~M_\odot$ have Eddington-limited
accretion. Therefore, we take the accretion mode of
$\lambda=2.0(M_{\rm{BH}}/2\times 10^9M_\odot)^{0.5}$
and reproduce the optical AGN LF at $z\sim 6$. The results are shown in Figures 2 and 3.
With this accretion mode, it is possible to take an AGN lifetime of $2\times 10^5$ yrs so
that our results are roughly consistent with the observational data of \citet{willott10b} and \citet{kashikawa14}. This
indicates that those AGNs at $z=6$ detected by \citet{willott10b} and \citet{kashikawa14} may be very young.

We also predict optical AGN LFs at $z=8$
and $z=9.5$ shown in Figure 3. Two parameter sets, one set for the
sub-Eddington accretion of $\lambda=0.25(M_{\rm{BH}}/2\times 10^9M_\odot)^{0.5}$
and an AGN lifetime of $2.0\times 10^7$ yrs, and the other set for the Eddington accretion of
$\lambda=2.0(M_{\rm{BH}}/2\times 10^9M_\odot)^{0.5}$ and an AGN lifetime
of $2.0\times 10^5$ yrs, are adopted in our prediction. The $M_{\rm{BH}}-\sigma$ relation
of \citet{robertson06} is used in the above calculations. These estimated results can be constrained
by future optical observations and surveys as well.

Two kinds of AGN LFs, one derived with the theoretical
$M_{\rm{BH}}-\sigma$ relation \citep{robertson06} and the other
with the observational $M_{\rm{BH}}-\sigma$ relation
\citep{gultekin09}, are given in this paper. We put lower and
upper limits on our AGN LFs by considering the intrinsic scatter of
the observational $M_{\rm{BH}}-\sigma$ relation.
We can distinguish different AGN LFs shown in Figures 1 and 2. Although our AGN LFs
derived from the theoretical $M_{\rm{BH}}-\sigma$ relation are
in good agreement with those derived from the observational
$M_{\rm{BH}}-\sigma$ relation, we clearly see that the AGN LFs at high redshifts have
uncertainties.

Both type-1 (broad-line)
and type-2 (narrow-line) AGNs are included in the calculation of the optical AGN LFs. The observational sample
adopted in this paper contains only type-1 AGNs. In principle, with X-ray telescopes,
type-1 AGNs are detected as unobscured AGNs, and most type-2 AGNs can be
effectively detected as obscured AGNs (Risaliti et al. 1999). However,
X-ray-selected AGNs and optical-selected AGNs have different classifications (Gilli et al. 2001;
Matt 2002). Because we have neither a one-to-one correspondence between
type-1 and unobscured AGNs, nor that between type-2 and obscured AGNs,
it is difficult to know the fraction of type-1 AGNs and type-2
AGNs in the optical band at high redshifts \citep{mayo13}.
Although the number density of obscured AGNs is four times larger than that
of unobscured AGNs (Risaliti et al. 1999), low-luminosity AGNs
are more frequently obscured than high-luminosity AGNs (Lawrence
1991; Ueda et al. 2003).

We obtain the AGN LFs in the $2-10$ keV band using the $M_{\rm{BH}}-\sigma$ relations of
\citet{robertson06} and \citet{gultekin09}. 
The AGN LFs at $z=2.25$, $z=3.0$, and $z=4.5$ are shown in Figures 4-6, respectively. 
A sub-Eddington accretion of 
$\lambda=0.25(M_{\rm{BH}}/2\times 10^9M_\odot)^{0.5}$
and an AGN lifetime of $9\times 10^7$ yrs are adopted in these
calculations. 
For comparison, we show the observational data from the CDF and the all-wavelength extended growth strip international survey (AEGIS) in the
redshift ranges of $2.0<z<2.5$ and $2.5<z<3.5$ \citep{aird10}. In the work of \citet{aird10}, the spectral identification and
the color pre-selected identification were used for AGN selection. In particular,
the color pre-selected processes are as follows. AGN candidates are first selected by
the Lyman-break technique \citep{steidel03} from the optical deep surveys based
on their optical colors. Then these color pre-selected objects are matched by
X-ray point sources. Our results show general agreement with the observed data. In Figure 6, the AGN LF at $4.0<z<5.0$ given by \citet{ueda14} is also shown. At the faint end, our results have excess compared to the observational data, but the overall trend agrees with most observational data points. 

Figure 7 shows our AGN LFs at redshifts 4.5, 5, and 6, respectively. 
The $M_{\rm{BH}}-\sigma$ relation of Robertson et al. (2006) is
adopted to reproduce these AGN LFs. The observational AGN LF at $4.0<z<5.0$ (Ueda et
al. 2014) is also shown. We see that the number of luminous AGNs
decreases with increasing redshift in our model. In contrast, at the low-luminosity part, the AGN LFs do
not show any flattening evidence. These results are consistent
with the trends for the observational AGN LFs given by Aird et al. (2010).
Although the hierarchical structure formation feature is shown in the X-ray
AGN LFs at $z>3$, large X-ray data samples are required to further identify
the AGN evolution properties at high redshifts.

\subsubsection{High-redshift AGN Number Counts in the X-Ray Band}
Some investigations of high-redshift AGN number counts in the X-ray band
are also important.
We follow the formalism of the cumulative number counts given by
\citet{ballantyne06} as follows:
\begin{equation}
N(>F)=\frac{K_{sr}c}{H_0}\int_{z_{min}}^{z_{max}}\int_{max(logL_{min},logL_F)}^{logL_{max}}
\Phi(L_x,z) 
\times\frac{d_l^2}{(1+z)^2[\Omega_m(1+z)^3+\Omega_\Lambda]^{1/2}}dlogL_xdz,
\end{equation}
where $\Phi(L_x,z)$ is the X-ray AGN LF, $K_{sr}=3\times 10^{-4}$ is the conversion factor from
$sr^{-1}$ to $\rm{deg}^{-2}$, $L_{\rm{F}}$ is the rest-frame
luminosity corresponding to the observed-frame flux $F$ at redshift
$z$, $d_l$ is the luminosity distance, $logL_{min}=41.5$ and
$logL_{max}=48$.

We take the observational AGN number counts given by both
\citet{brusa09} from the XMM-COSMOS survey and \citet{civano11} from the 
{\it{Chandra}}-COSMOS sample to compare with our results.
AGN candidates from the XMM-COSMOS survey have spectral identifications in the optical band,
and optical broad-lines are shown in the optical spectrum of each
AGN. Half of the AGN sources in the {\it{Chandra}}-COSMOS sample also have
optical broad-lines shown in their optical spectra. Therefore, we
treat these X-ray selected AGNs as unobscured sources. However,
Equation (2) contains contributions from both unobscured and
obscured AGNs.

To specify the contribution of the unobscured/obscured AGNs to the
AGN number counts in the X-ray band so that we can compare our
results with those of the observational data, we use the statistical
results given by \citet{hasinger08}. In that work, the obscured
AGN fraction decreases with the X-ray luminosity but increases with
redshift as $f=-0.226logL_x+10.306$ and $f=0.308(1+z)^{0.48}$,
where $f$ is the obscured AGN fraction. In this subsection, it is essential that
we apply the intrinsically absorbed X-ray spectrum to obtain the observational
X-ray flux. Then, we can derive the AGN number counts in the X-ray band. Given a certain intrinsic
absorption $\rm{N_H}$ value, the X-ray spectral template
can be confirmed. In Appendix A, we put some examples of intrinsically absorbed
X-ray spectra at high redshifts.
Finally, we can obtain high-redshift AGN number counts in the X-ray band with certain
$\rm{N_H}$ values. Here, we notice one
difference between the obscuration mentioned in this subsection
and the attenuation from the AGN host galaxy. Based on the AGN unified
model \citep{antonucci03}, the classification of
unobscured/obscured AGNs can be defined through X-ray spectra and
X-ray hardness ratios. The attenuation property is decided by the 
AGN torus with different view angles, while the attenuation of
an AGN host galaxy is identified as the absorption caused by dusty
gas surrounding the central BH. However, the dusy gas is also the material
for both central BH growth/accretion and star formation. We also note
that \citet{hopkins11} combined two attenuation origins as one
effect from the AGN torus.

We show the AGN number counts in the $0.5-2$ keV
band at $z>3$ and $z>4$ in Figures 8 and 9, respectively. The AGN number counts with different absorption
$\rm{N_H}$ values
($1.0\times 10^{22}~\rm{cm^{-2}},~5.0\times 10^{22}~\rm{cm^{-2}},~and~
1.0\times 10^{23}\rm{cm^{-2}}$) are given in each figure.
The theoretical $M_{\rm{BH}}-\sigma$ relation \citep{robertson06} is considered.
Although our results of the X-ray AGN number counts are consistent with the
observational data in general, the uncertainties of these results from the X-ray intrinsic attenuation are clearly shown.

We show the AGN number counts in the
$0.5-2$ keV band at $z>4$, $z>5$, and $z>6$, respectively, in Figure 10.
Our results derived from the $M_{\rm{BH}}-\sigma$ relation of
\citet{robertson06} are shown. The absorption $\rm{N_H}$ is fixed as
$5.0\times 10^{22}~\rm{cm^{-2}}$. We also show the observational
AGN number counts at $z>4$ given by \citet{brusa09} and \citet{civano11} in
Figure 10.
\citet{gilli11} discovered one Compton-thick AGN at redshift 4.76 in the 4 Ms CDF-S field. This finding provides
some observational evidences of the AGN
attenuation properties: a hardness ratio of 0.23 and an intrinsic
column density of $10^{24}~\rm{cm^{-2}}$
indicating a heavy absorption in the X-ray band, and
a multi-wavelength SED fitting suggesting a strong star formation
process. Compton-thick AGN surveys have recently been performed \citep{brightman14,lanzuisi14}.
We hope that our results for the high-redshift AGN population
can be constrained by all kinds of updated X-ray detections.

\section{Simple Estimation of the AGN Contribution to Cosmic Reionization}
The cosmic reionization epochs at redshift 6 and redshift 10.9 were
constrained by the Gun-Peterson trough \citep{gunn65} of the 
high-redshift quasar spectra \citep{fan02,fan06} and the cosmic
microwave background polarization of the Wilkinson Microwave
Anisotropy Probe (\emph{WMAP}) measurements \citep{spergel06},
respectively. Two fundamental points are involved in the topic of
the cosmic reionization: (1) what kind of object can emit ionizing
photons for the cosmic reionization and (2) how many ionizing photons can
be emitted by these objects? It was suggested that some ionizing
photons emitted from high-redshift AGNs can survive after 
attenuation by the surrounding medium; then, these photons can ionize the IGM
\citep{madau99}. In this section, we estimate the contribution of
high-redshift AGNs to the cosmic reionization.

We have calculated the optical AGN LFs at high
redshifts in Section 2.2. Thus, we can sum up AGN emissions
at a certain redshift and obtain the total ionization rate as
$\dot{N}_{\rm{ion}}=\int L\Phi(L,z)/(h\nu)dL$, where $h$ is the Planck constant,
and $\nu$ is the average ionizing frequency \citep{choudhury05}. 

To quantify the cosmic reionizing process of neutral
hydrogen, the time evolution of the filling factor $Q_{\rm{HII}}$ has been
derived by \citet{shapiro87} as follows:
\begin{equation}
\frac{dQ_{\rm{HII}}}{dt}=\frac{\dot{N}_{\rm{ion}}}{n_{\rm{H}}}-\frac{Q_{\rm{HII}}}{t_{\rm{rec}}},
\end{equation}
where $n_{\rm{H}}=1.7\times 10^{-7}~\rm{cm^{-3}}$ is the number
density of neutral hydrogen in the local universe,
$t_{\rm{rec}}=0.3((1+z)/4)^{-3}(C/30)^{-1}$ Gyr is the recombination
time, and $C$ is the clumping factor. Different solutions for
Equation (3) were fully discussed by \citet{meiksin05}.
$Q_{\rm{HII}}\ge 1$ indicates the fully ionized case of the cosmic
reionization, and the corresponding redshift is the reionization
epoch. The clumping factor
$C(z)=17.6e^{-0.1z+0.0011z^2}$ was reported by \citet{iliev06}. In the
relevant reionization redshift interval of $7<z<10$, we obtain
$7<C<10$. Here we adopt $C=7$ as a reference value \citep{kaurov14}. The photon number contributed to the cosmic reionization $N_{\rm{ion}}$ is multiplied by an escape fraction $f_{\rm{esc}}$, because a part of photons emitted from AGNs are absorbed by the surrounding medium of AGNs and the AGN host galaxies. It is quite complicated
to determine the value of the escape fraction $f_{\rm{esc}}$. A brief
discussion was given by \citet{barton04}. It was pointed out that this parameter is dependent on star formation
processes and dust production in primordial galaxies \citep{mao07}. Here, we
take $f_{\rm{esc}}=0.5$ as a reference value.

We obtain different evolution traces for the filling factor. The results are shown in Figure 11. AGNs with a BH mass range of $1.0\times 10^6-1.0\times 10^7~M_\odot$ can reionize the Universe at $z\sim 8$, while AGNs with a BH mass range of $1.0\times10^7-1.0\times 10^8~M_\odot$ can also reionize the Universe at $z\sim 7$. However, AGNs with a BH mass range of $1.0\times 10^8-1.0\times 10^9~M_\odot$ cannot fully reionize the Universe at $z>6$. Therefore, we conclude that low-luminosity AGNs dominate the cosmic reionization. Recently, Bouwens et al. (2015) estimated some cosmic reionization processes. Using the SDSS AGN LF at $z=5$ given by McGreer et al. (2013), they built luminosity evolution and density evolution models. They claimed that AGNs do not have enough photons for the cosmic reionization at $z>6$. However, Giallongo et al. (2015) analyzed the AGN ionization emissivity at $z>4$. Some faint AGN candidates at $z=4-6.5$ have been detected from the GOODS observation, and optical AGN LFs have been derived. It has been suggested that the AGN population provides an important contribution to the cosmic reionization. From a theoretical point of view, Volonteri \& Gnedin (2009) presented a simple high-redshift BH growth model. They also noted the importance of the AGN population for the cosmic reionization. Recently, Hao et al. (2015) proposed that intermediate BHs (mass range of $10^2-10^5~M_\odot$) shining at high redshifts emit UV photons to ionize the Universe. In our model, we predict the optical AGN LFs at $z=6$. We estimate that photons from high-redshift AGNs can effectively ionize the universe at $z>6$. However, at $z=6$, the optical LF at the faint end in our model is 10 times that of the observational one. We expect that more faint AGNs at $z>6$ can be detected in the future. 

Although high-redshift AGNs seem to provide enough photons
to fully reionize the Universe beyond redshift 6, the following considerations should be taken into account.
First, those optical AGN LFs shown in Figure 2 have large
uncertainties at the faint end. In particular, at a magnitude of
$M_{1450}=-22.2$, our AGN LF at $z=6$ is 10
times larger than the observational one. Second, the determined
reionization epochs in Figure 11 are dependent on the clumping
factor $C$. Finally, the escape fraction $f_{\rm{esc}}$ value is referenced
from the environment of young galaxies at high redshifts. We assume that the escape fraction of AGNs is the same as that of
high-redshift galaxies. In summary, we expect that more faint AGNs beyond redshift 6 can be identified in the future
if the cosmic reionization can indeed be dominated by the AGN activity.

Some of the primeval galaxies at redshift $6-10$ detected from optical
deep surveys are also candidates to fully reionize the Universe \citep{bouwens06,bunker10,robertson10,bouwens11}. Some
reionization processes have been fully discussed
\citep{dijkstra07, mao07, salvaterra10}. It is known that different kinds of objects are involved in
cosmic reionization processes.
Within the framework of joint evolution between
central BHs and their host galaxies, we suggest two reionization stages. Ionizing
photons emitted from proto-galaxies dominate the first stage
of the cosmic reionization. In this stage, central BHs with masses
less than $10^6$ $M_\odot$ are hidden inside proto-galaxies
and ionizing photons are fully absorbed
by the surrounding medium.
After sufficient central BH growth and star formation,
central BHs shine as typical AGNs, and the AGN environment is
transparent. Ionizing photons emitted from AGNs may
dominate the second stage of the cosmic reionization.

\section{Discussion}

\subsection{``Flicker" Accretion}
Accretion is one of the key points for AGN shining. As mentioned
in Section 1, with the sub-Eddington accretion of $\lambda\sim 0.5$,
\citet{shankar09} overestimated the number density of faint AGNs
in the luminosity range of $-25.5<M_{1450}<-24.5$ at $z=6$. 
We know that the QSOs at $z=6$ found by
\citet{willott10,willott10b} and \citet{kashikawa14} have an Eddington ratio close to the Eddington limit.
In this paper, if we increase the Eddington ratio
to the Eddington limit and decrease the AGN lifetime to $2\times 10^5$ yrs, we can reproduce
the optical AGN LF at $z=6$.
This indicates that BHs with a mass of $10^8~\rm{M_\odot}$ accreting in the Eddington limit
at $z=6$ can form very young QSOs.
Although we use a single Eddington ratio per BH mass, we realize that the accretion history during BH growth is related to the Eddington ratio (Granato et al. 2004). Shen (2009) investigated the BH growth in detail taking into consideration the evolution of the Eddington ratio. Aird et al. (2012) reported an observational distribution of the Eddington ratio, which  is an universal function with a power-law shape. Thus, rather than a single Eddington ratio per BH mass, the Eddington ratio distribution should be taken into account when modeling the AGN LFs. In our model, we use the single Eddington ratio form without any distribution function. We ignore complicated BH accretion processes during BH growth, and we think that the single Eddingtion accretion mode corresponds to the peak of the AGN lightcurve. Furthermore, because we find the AGN ``flicker" phase at high redshift, it is reasonable that the peak of the AGN light curve occurs over the time of the AGN “flicker” phase. This peak/``flicker" time is much shorter than the averaged AGN lifetime.  Therefore, the single accretion mode without any distribution scatter is helpful to explore the AGN short-time variability/burst given by Novak et al. (2011) and Schawinski et al. (2015).

\subsection{AGN Lifetime}
We propose a possible lifetime of $2.0\times 10^5$ yrs for the optical AGNs at $z\ge 6$, and these high-redhsift optical AGNs are younger than those in previous studies. For example, Bolton et al. (2011) provided the observational AGN bright phase at $z>6$, and the AGN lifetime was constrained as $1.0\times 10^6-1.0\times 10^7$ yrs. Here, we further summarize some physical
possibilities regarding on the AGN lifetime. There have been more high-redshift ($z\ge 6$) AGN findings (Reed et al. 2015; Venemans et al. 2015; Wu et al. 2015). If we use the BH growth scenario to explain massive AGNs at high-redshift, we expect massive AGN activity within the timescale of $1.0\times 10^7-1.0\times 10^8$ yrs. The AGN e-folding time and duty circle should be considered in the cosmological evolution scenario. However, it is still possible to have short-timescale variabilities in the AGN luminosity during the BH growth phase. This possibility could particularly be important for the study of high-redshift AGNs. Novak et al. (2011) performed two-dimensional simulations to study the time-dependence of BH accretion, and the AGN light curve within an AGN phase of 250 Myr can be divided by many short bursts with a typical timescale of $1.0\times 10^5$ yrs. Gas supply for BH growth can be through instability-driven inflows (Bournaud et al. 2011). If BH growth occurs in a star-forming disk galaxy, BH accretion with short-time variability has  been identified in numerical simulations (Gabor \& Bournaud, 2013). Hickox et al. (2014) studied the connection between the BH activity and the star formation. They proposed that the timescale of 100 Myr corresponds to an average number of the whole BH accretion, while instantaneous luminosity
corresponds to a variability with a much shorter timescale. Very recently, Schawinski et al. (2015) observationally defined the AGN lifetime, which is the ratio between the time to photoionize a host galaxy and the fraction of optically elusive AGNs. Due to the arguments on the switch-on (high-accretion rate) and switch-off (low-accretion rate) processes, Schawinski et al. (2015) claimed that the AGN ``flicker" duration is $1.0\times 10^5$ yrs based on the {\it Swift}-BAT and SDSS samples. We think that their results can support an AGN lifetime number on the order of $10^5$ years adopted in our model for those Eddington-limited AGNs at $z\ge 6$. In our model, the AGN lifetimes at different redshifts have different numbers. The AGN lifetime of $2.0\times 10^7$ yrs is used in our model to reproduce the AGN LF at $z=4$. If we take the ``flicker" scenario given by Schawinski et al. (2015), we think that the AGN lifetime with an order of $10^7$ yrs indicates an averaged AGN activity process with sub-Eddington
accretion. 
At $z\ge 6$, we do not have enough observational data, and the observational AGN LF spreads in a relatively small range of luminosity. If we reproduce the optical AGN LF using the Eddington-limited accretion at $z\ge 6$, the ``flicker" AGN lifetime of $2.0\times 10^5$ yrs in the BH growth scenario can be understood as the time of the instantaneous AGN activity. We also speculate (1) there are enough gas clumps surrounding the central BH to fuel AGN shining at high redshift, so that the accretion rate is high, and we could identify violent short-time bursts of
AGNs, and (2) at low-redshift, there are fewer gas clumps surrounding the central BH, so that the accretion rate is low, and we
observe many AGNs without any violent short-time bursts. Thus, we emphasize that the general AGN lifetime
is much longer than the lifetime of short-time bursts.

\subsection{Feedback Effects}
The AGN/supernova kinetic feedback effects are strongly involved
in the galaxy formation and evolution: (1)  heating the surrounding
gas of a central BH and quenching the star formation of a galaxy bulge and
(2) blowing away the surrounding gas of a central
BH and stopping further BH growth. In particular, we note that the AGN
feedback dominates in massive dark matter halos while the supernova
feedback is effective in small dark matter halos
\citep{granato04}. We further discuss the
feedback effect of supernovae on the faint end of the AGN LF mentioned by
\citet{lapi06} and \citet{fontanot07}: a small dark matter halo
potential well is not deep enough
to contain cooling baryons if we consider supernova feedback, and further BH growth is not
expected. Hence, there are no more low-mass BHs in small dark
matter halos, and AGN LFs show a flattening feature at the
faint end. However, this reasonable explanation is challenged by
the optical AGN LF at $z\sim 4$ \citep{glikman10}. The AGN LF
given by \citet{glikman10} does not show the flattening feature,
indicating more faint AGNs in the high-redshift
universe\footnote{A shallow slope of the optical LF at $z\sim 4$ was
reported by \citet{ikeda11}. However, we note that the measurement of
\citet{glikman10} is 2-3 mag deeper than that of
\citet{ikeda11}.}. Although it is likely that the supernova
feedback effect on AGN LFs reported by
\citet{lapi06} and \citet{fontanot07} is overestimated, we note that Masters et al. (2012) also 
presented the flattening feature at the faint end of the AGN LF. Moreover,
strong supernova feedback can blow away the material surrounding
a central BH, while the material surrounding a central BH may exist
when the supernova feedback is weak. Thus, weak supernova feedback
can also be one possible reason that
low-luminosity AGNs harbored in small dark matter halos suffer strong attenuation. In order
to obtain quantitative results, some star formation and evolution
processes, including the initial mass function and the star
formation rate, should be explored in the future.

\subsection{BH Merger Rates}
Although we assume that BH accretion dominates the AGN
evolution process, BH merger effects should be estimated.
We assume that one BH is inside one dark matter halo, and
the dark matter halo merger is followed by the central BH merger. Then, we can
apply a simple treatment reported by \citet{shankar09}. The
fraction of the BH merger is given as $F=P_{merg}\Delta
t/t_{\rm{H}}(z)$, where $P_{merg}$ is the probability of the BH
merging with equal-mass BH, $t_{\rm{H}}(z)$ is the Hubble time,
and $\Delta t$ is the dark matter halo merging timescale. We
assume $P_{merg}=0.5$. From an investigation of the BH/dark matter
halo merger \citep{kulkarni11}, we can obtain the value of $\Delta
t/t_H(z)$. The variation of the BH mass function is $\Delta
\Phi=F\times \Phi(M_{\rm{BH}}/2)/2-F\times \Phi(M_{\rm{BH}})$.
Finally, we obtain the BH merger effect on AGN LFs. As an example, in the case in which the dark matter halo has a mass of
$1.0\times 10^{12}~M_\odot$, the BH merger effect provides
variations of 17\% and 15\% on the AGN LFs at $z=4$ and $z=6$,
respectively. Moreover, the variation of BH accretion caused by the 
BH merger can be ignored \citep{shankar10}. Besides these simple
estimations mentioned above, in the following, we discuss some
complicated issues. The so-called ``dry merger'', which refers to the dark
matter halo merger without baryons, has been put forward
\citep{vandokkum05}. In this case, BHs do not merge even if dark
matter halos have merging processes. This is one reason that not
all of the major mergers end up with massive early-type galaxies
\citep{chou11}. On the other hand, the dark matter halo merging
timescale and the incoming BH merging timescale were studied by
\citet{shen09} and \citet{kulkarni11}. At high-redshift, BHs
sink to the center of the merged dark matter halo before BH
merging. Therefore, at high redshift, binary BHs or triple BHs can
stay in the same host galaxy. Thus, the $M_{\rm{BH}}-\sigma$
relation should be revisited. These interesting topics
are the next step in the future research.

\subsection{High-redsift AGN Number Counts}
The predictions of high-redshift AGN detection are
helpful for future X-ray mission proposals.
The WFXT deep survey can
cover an observational area of 100 $\rm{deg}^2$ (Rosati et al. 2010). In
our model, to the flux limit of $3\times
10^{-17}~\rm{erg~cm^{-2}~s^{-1}}$ in the $0.5-2$ keV band, we
predict that $\sim$30 AGNs per $\rm{deg}^2$
with the absorption of $\rm{N_H}=5.0\times 10^{22}~\rm{cm^{-2}}$ can be
detected at $z>6$. 
On the other hand, high-redshift
AGN detections have been investigated by other theoretical models
as well.
At $z>6$, to the flux limit of $3\times 10^{-17}~\rm{erg~cm^{-2}~s^{-1}}$ in the
$0.2-2$ keV band, optimistic detection of $500-600$
AGNs per $\rm{deg}^2$ and pessimistic detection of $\sim$ 16
AGNs per $\rm{deg}^2$ were predicted by \citet{rhook08} and
\citet{marulli08}, respectively. Although Marulli et al. (2008) successfully predicted the general BH mass function, they underestimated the number density of high-z bright AGNs when they used a constant Eddington accretion.  While Rhook \& Haehnelt (2008) extrapolated their merger-driven accretion model to high redshift for BH growth scenarios, and the predicted AGN numbers were sensitive on the early BH growth history. In our model, the BH accretion is adjusted by the BH mass.
The value predicted in our model is two times larger than
the value of \citet{marulli08}, but $16-20$ times smaller than the
value of \citet{rhook08}. All of these high-redshift
predictions can be well verified by future observations.

\subsection{Brief Illustration of Model Methodology}
We use a simple analytical model to study the high-redshift AGN evolution in this paper. This model adopts some physical relations, such as 
$M_{\rm{BH}}-\sigma$ and $\sigma-V_c$ relations, which are confirmed by observations and simulations. Because this is a phenomenological model, as discussed in previous subsections, some important physical properties related to the high-redshift AGN evolution, such as star formation, AGN/SN feedback, and BH accretion/merger, are not incorporated. However, semi-analytical models and hydro-dynamical simulations can self-consistently describe star formation, BH accretion, radiative cooling, and SN/AGN feedback processes under the framework of co-evolution between central BHs and their host galaxies.

In order to obtain enough spacial and mass resolutions of N-body simulations, especially for studying high redshift ($z>7$) AGN evolutions, huge simulation box sizes and a large number of simulation particles inside each box are required. Hydro-dynamical simulations are also limited by the calculation grids in their computational domain. In the case of semi-analytical models, it usually takes heavy computational time to decide parameters for obtaining reasonable results. However, the calculations in our model have no limits on the box sizes, number of input particles, grids of the computational domain, and input parameters. This kind of phenomenological model can be applied as a preliminary test for semi-analytical models to comprehensively study the high-redshift AGN evolution. Furthermore, some interesting issues revealed in our model for the high-redshift AGN LFs, such as the AGN flicker accretion and the AGN short lifetime, can be also noted and examined in purpose by semi-analytical models.

\section{Conclusions}
To investigate the evolution of AGNs at high redshift, we build a simple model to
study high-redshift AGN LFs and AGN number counts. 
We have determined the following conclusions:
\begin{enumerate}
\item With a moderate sub-Eddington ratio and an AGN lifetime of about $10^7$-$10^8$ years, optical and X-ray AGN LFs at high redshifts are produced.
At $z\sim 4$, we do not obtain any flattening feature at the faint end of the optical
AGN LF, consistent with observational results.
\item The optical AGN LF at $z=6$ built by \citet{willott10,willott10b} and \citet{kashikawa14} can be reproduced with Eddington-limited
accretion and an AGN lifetime of about $2\times 10^5$ yrs. It seems possible that these AGNs at $z=6$ are very young.
\item We calculate the X-ray AGN LFs at $2.0<z<5.0$ and the X-ray AGN number counts at
$z>3$ using the same parameters adopted for the calculation of the optical AGN LFs at $z\sim 4$.
We estimate that, at $z>6$, there are about 30 AGNs per $\rm{deg}^2$ to the
flux limit of $3\times 10^{-17}~\rm{erg~cm^{-2}~s^{-1}}$ in the $0.5-2$ keV band.
\item High-redshift AGNs with central masses less than $10^8~M_{\odot}$ may dominate the cosmic
reionization, though
this domination is strongly dependent on the AGN LF, clumping factor, and escape fraction of ionizing photons.
\end{enumerate}

Although our AGN LFs and AGN number counts at high redshifts have uncertainties that originate from the
$M_{\rm{BH}}-\sigma$ relation, Eddington ratio, AGN lifetime, and X-ray attenuation, they can be further 
constrained by optical and X-ray observations (Subaru/PFS, WFXT, eROSITA) in the future.



\acknowledgments

The observational data are kindly provided by J. Aird, A. Bongiorno,
M. Brusa, F. Fontanot, E. Glikman, and L. Jiang. J. M. is
grateful to X.-B. Wu, Y. Lu, W. Zuo, F. Fontanot, and Y. Ueda for their useful
discussion. J.M. was encouraged by Shigehiro Nagataki for this work. J. M. is supported by the Hundred Talent Program of Chinese Academy of Sciences, the Major Project Program of Chinese Academy of Sciences (KJZD-EW-M06), and the Introducing Oversea Talent Plan of Yunnan Province. M. K. was supported by Basic Science Research Program
through the National Research Foundation (NRF) of Korea funded by
the Ministry of Education, Science and Technology (2012-0001787).

\appendix

\section{Absorbed X-ray Spectral Templates}
The observed AGN number counts in the X-ray band given by \citet{brusa09} and \citet{civano11} are related to the observed X-ray flux. Both the absorption $N_{\rm{H}}$ and the flux density of the X-ray spectrum in a given redshift should be determined to transform from intrinsic number counts to observed number counts.

In order to study the absorption feature of X-ray selected AGNs and obtain high-redshift AGN number counts in the  
X-ray band, we should investigate X-ray spectral templates at high redshifts. The model with command ``$wabs*zwabs*pow$''
can be used in XSPEC (version 12), as we assume that the spectrum can be fitted by an absorbed power-law. The Galactic absorption is taken as $\rm{N_H}=2.9\times 10^{20}~\rm{cm^{-2}}$
\citep{dickey90}. We assume a photon index $\Gamma=2.0$ \citep{gilli07} in the absorbed power-law spectrum. We show the X-ray
spectra at redshifts 3, 4, 5, and 6 in Figure 12. Meanwhile,
the spectra with the intrinsic absorption values of $\rm{N_H}=2.5\times 10^{20},~5.0\times 10^{21},~
5.0\times 10^{22},~5.0\times 10^{23},~\rm{and}~2.5\times 10^{24}~\rm{cm^{-2}}$ are also given.

The normalized fluxes in the $0.5-2$ keV band with a certain $N_{\rm{H}}$ can be obtained from the above spectral fitting. Using these spectral results, we quantitatively calculate the absorbed X-ray fluxes at high redshifts. Finally, as described in
Section 2.2.2, the observed AGN number counts in the X-ray band can be derived.

\section{BH$-$Halo Mass Relation and BH$-$Baryon Mass Relation}
Although we focus on high-redshift AGN LFs and AGN number counts, we pay attention to the ratio of BH mass to dark matter halo mass and the ratio of BH mass to baryonic mass under the framework of joint evolution between AGNs and their host galaxies. We transfer dark matter halo mass to baryonic mass by a factor of 0.16. We show the plots of the BH$-$halo mass relation and the BH-baryon mass relation in Figures 13 and 14, respectively. In each Figure, the cases of redshift 4, 6, 8, and 9.5 are given. The different forms of the $M_{\rm{BH}}-\sigma$ relation and the $M_{\rm{BH}}-\sigma$ relation scatters are adopted in the calculations.

\clearpage



\begin{figure}
\includegraphics[angle=270,scale=0.65]{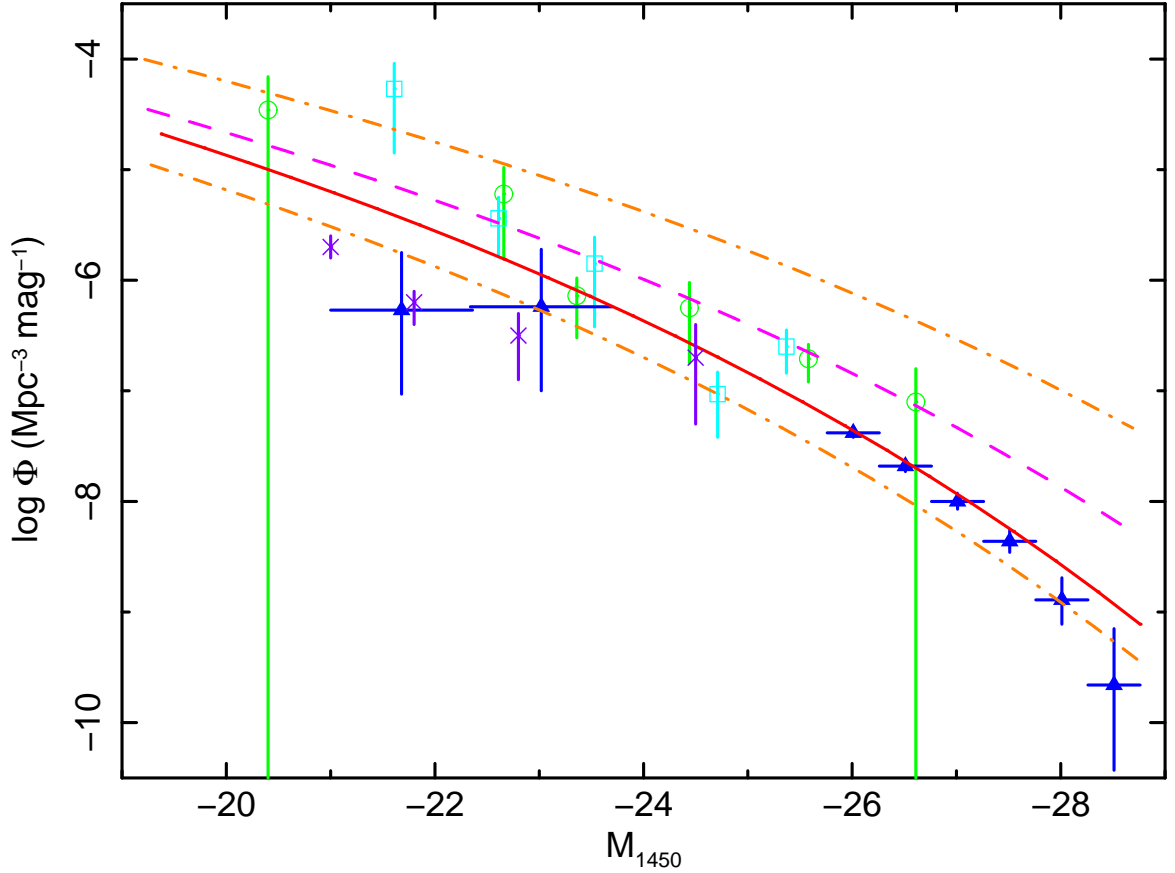}
\vspace{10pt}
\caption{Optical AGN LFs at $z=4$. Data points with error bars are
collected from \citet{fontanot07} (solid triangles),  \citet{masters12} (solid cross), and
\citet{glikman10} (photometric-redshift selection results denoted
by empty circles and spectral-redshift identification results
denoted by empty squares). The solid line denotes our
result, adopting the $M_{\rm{BH}}-\sigma$ relation from
numerical simulations given by \citet{robertson06}. The dashed line
denotes our result as well, adopting the observational
$M_{\rm{BH}}-\sigma$ relation given by \citet{gultekin09}. Two
dotted-dashed lines denote the lower and upper limits of our model
results in the case where the intrinsic scatter of the
observational $M_{\rm{BH}}-\sigma$ relation given by
\citet{gultekin09} is considered.}
\end{figure}

\begin{figure}
\includegraphics[angle=270,scale=0.65]{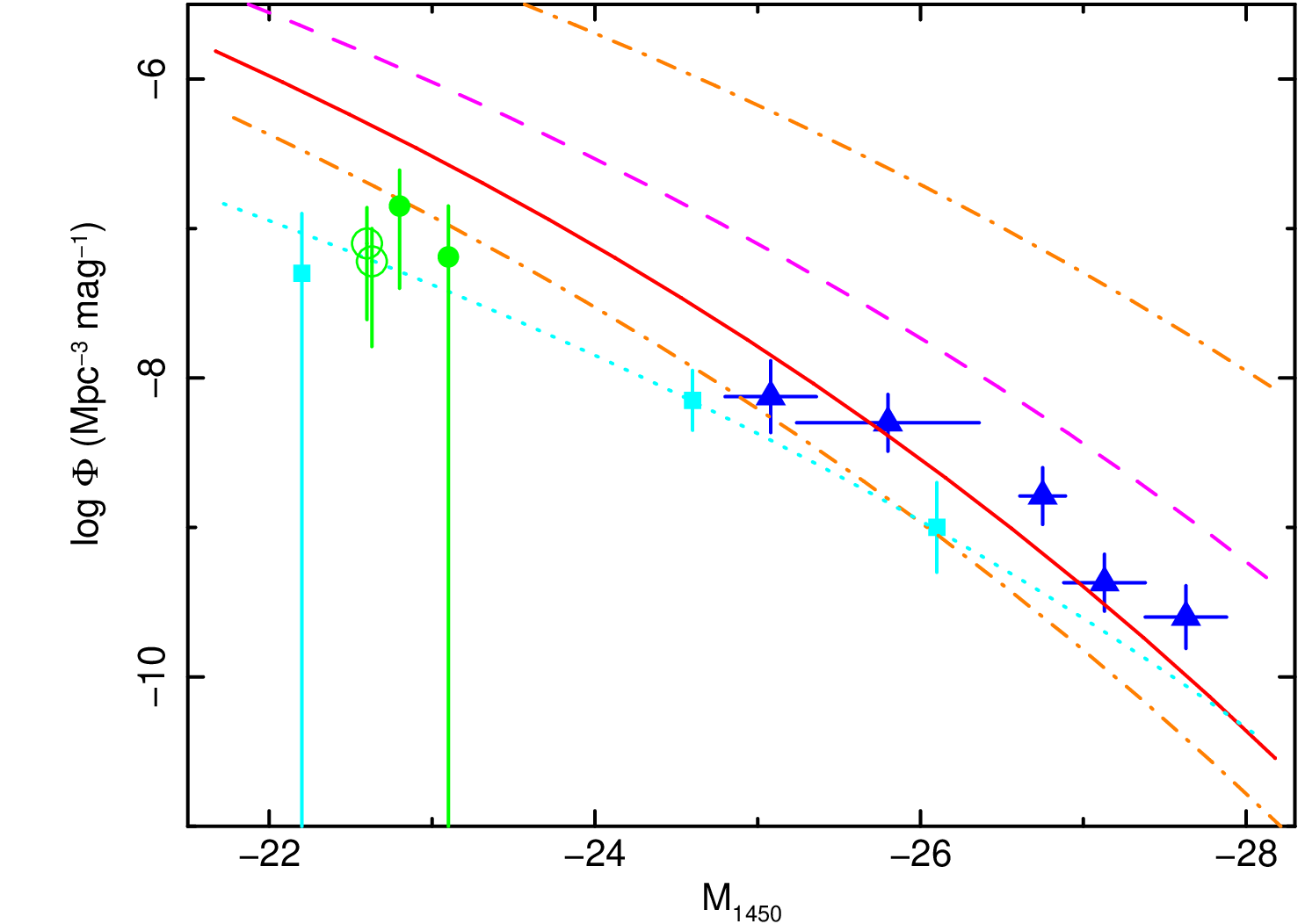}
\caption{Optical AGN LFs at $z=6$. Data points with error bars are
collected from \citet{jiang09} (solid triangles),
\citet{willott10b} (solid squares), and \citet{kashikawa14} (solid and empty circles).
Those denotations of solid,
dashed, and two dotted-dashed lines are the same as those in Figure 1.
In particular, the dotted line indicates the result of the calculation with the Eddington ratio of
$\lambda=2.0(M_{\rm{BH}}/2\times 10^9M_\odot)^{0.5}$ and a very short AGN lifetime of $2\times 10^5$ years. In this case,
the $M_{\rm{BH}}-\sigma$ relation given by \citet{robertson06} is adopted.
\label{fig2}}
\end{figure}

\begin{figure}
\includegraphics[angle=270,scale=0.65]{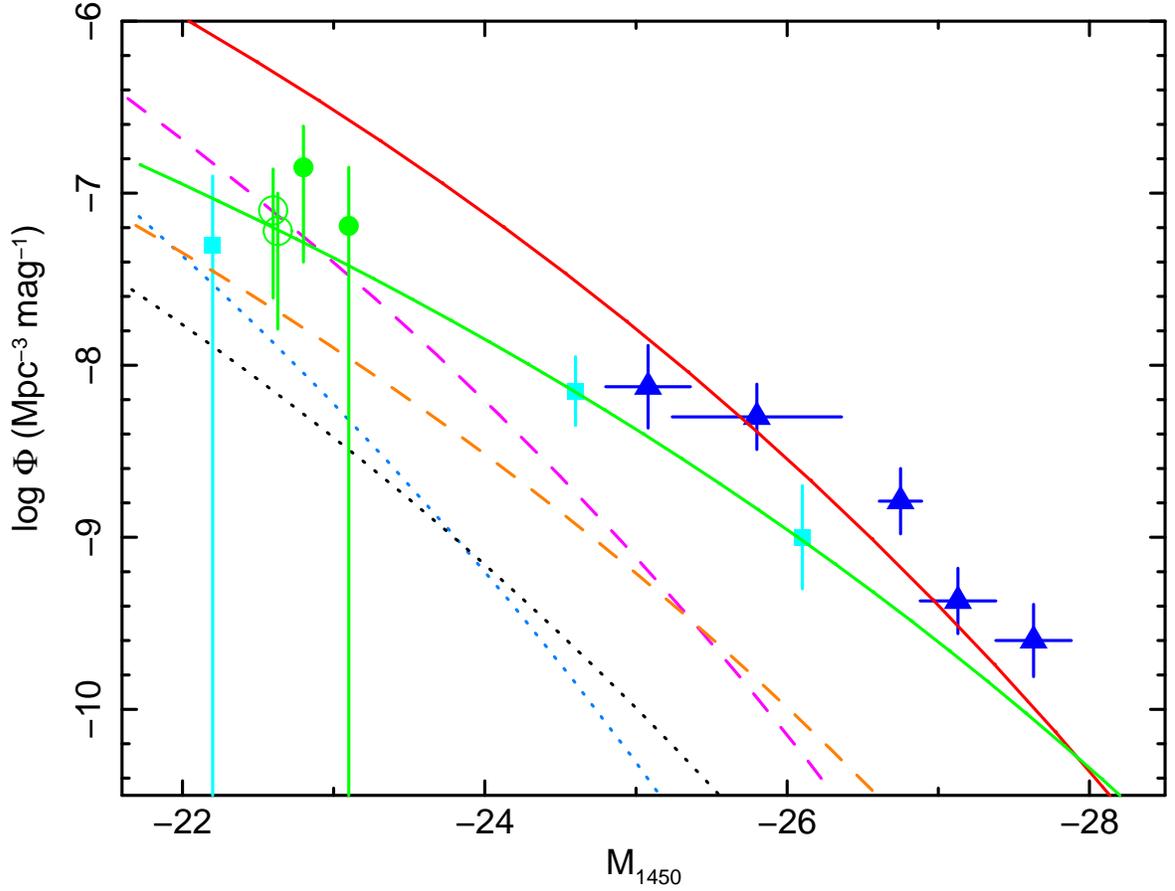}
\caption{Optical AGN LFs at $z=6$ (solid lines), $z=8$ (dashed lines), and $z=9.5$ (dotted lines),
respectively. Red, pink, and blue lines indicate the results calculated from the sub-Eddington
accretion of $\lambda=0.25(M_{\rm{BH}}/2\times 10^9M_\odot)^{0.5}$
and the AGN lifetime of $2\times 10^7$ years. Green, orange, and black lines indicate the results
calculated from the Eddington accretion of $\lambda=2.0(M_{\rm{BH}}/2\times 10^9M_\odot)^{0.5}$
and the AGN lifetime of $2\times 10^5$ years.
The $M_{\rm{BH}}-\sigma$ relation from the numerical
simulations given by \citet{robertson06} is used to reproduce all of these AGN LFs.
Data points with error bars are the same as those in Figure 2.
\label{fig3}}
\end{figure}

\begin{figure}
\includegraphics[angle=270,scale=0.65]{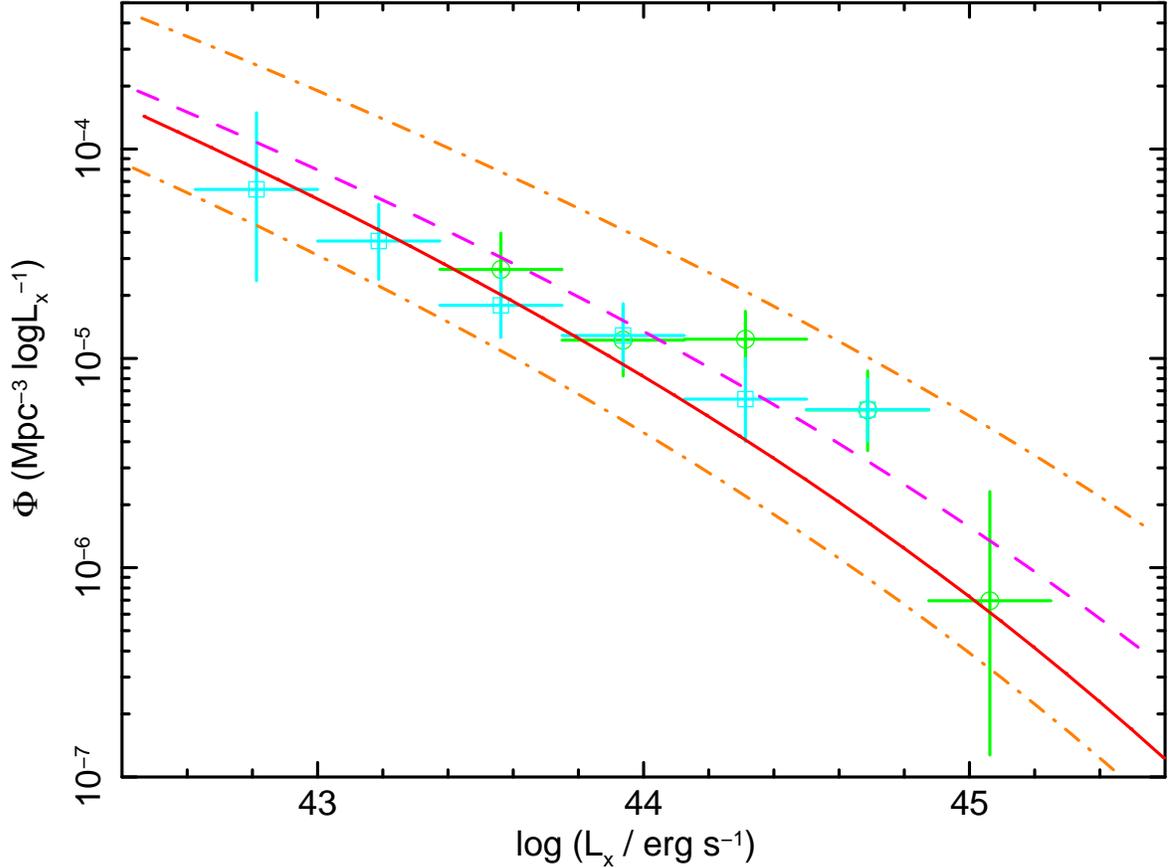}
\caption{AGN LFs in the $2-10$ keV band at $z=2.25$. Data points
with error bars are the observational results at
$2.0<z<2.5$ given by \citet{aird10}: open circles denote
the AGN LF obtained with the X-ray selected sample while open
squares denote the AGN LF obtained with the high-redshift color
pre-selected sample. 
The solid line denotes our result,
adopting the $M_{\rm{BH}}-\sigma$ relation from the numerical
simulations given by \citet{robertson06}. The dashed line denotes
our result as well, adopting the observational $M_{\rm{BH}}-\sigma$
relation given by
\citet{gultekin09}. Two dotted-dashed lines denote the lower and
upper limits of our results in the case that the intrinsic
scatter of the observational $M_{\rm{BH}}-\sigma$ relation given
by \citet{gultekin09} is considered. \label{fig4}}
\end{figure}

\begin{figure}
\includegraphics[angle=270,scale=0.65]{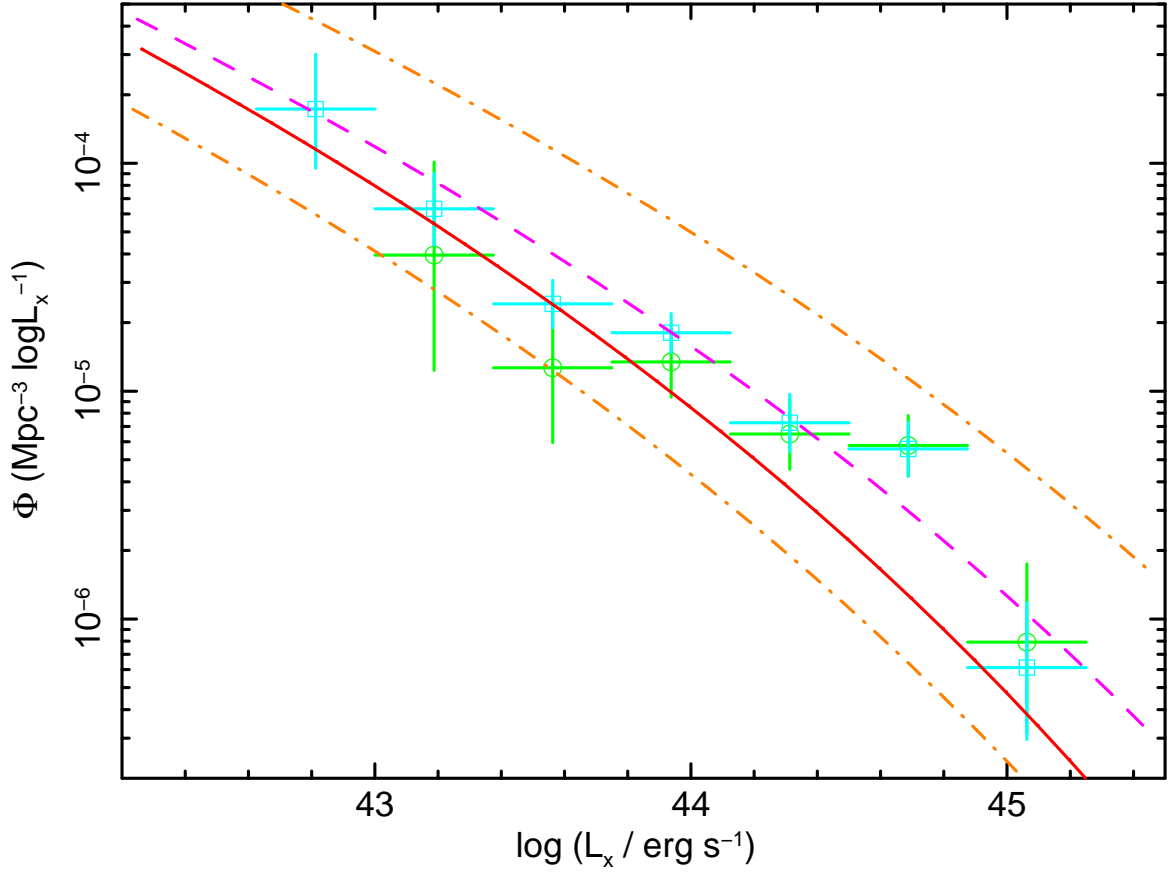}
\caption{AGN LFs in the $2-10$ keV band at $z=3.0$. The data
points with error bars are the observational results at
$2.5<z<3.5$ given by \citet{aird10}: the open circles
denote the AGN LF obtained with the X-ray selected sample while
the open squares denote the AGN LF obtained with the high-redshift
color pre-selected sample. The denotations of solid, dashed, and
two dot-dashed lines are the same as those in Figure 4.
\label{fig5}}
\end{figure}

\begin{figure}
\includegraphics[angle=270,scale=0.65]{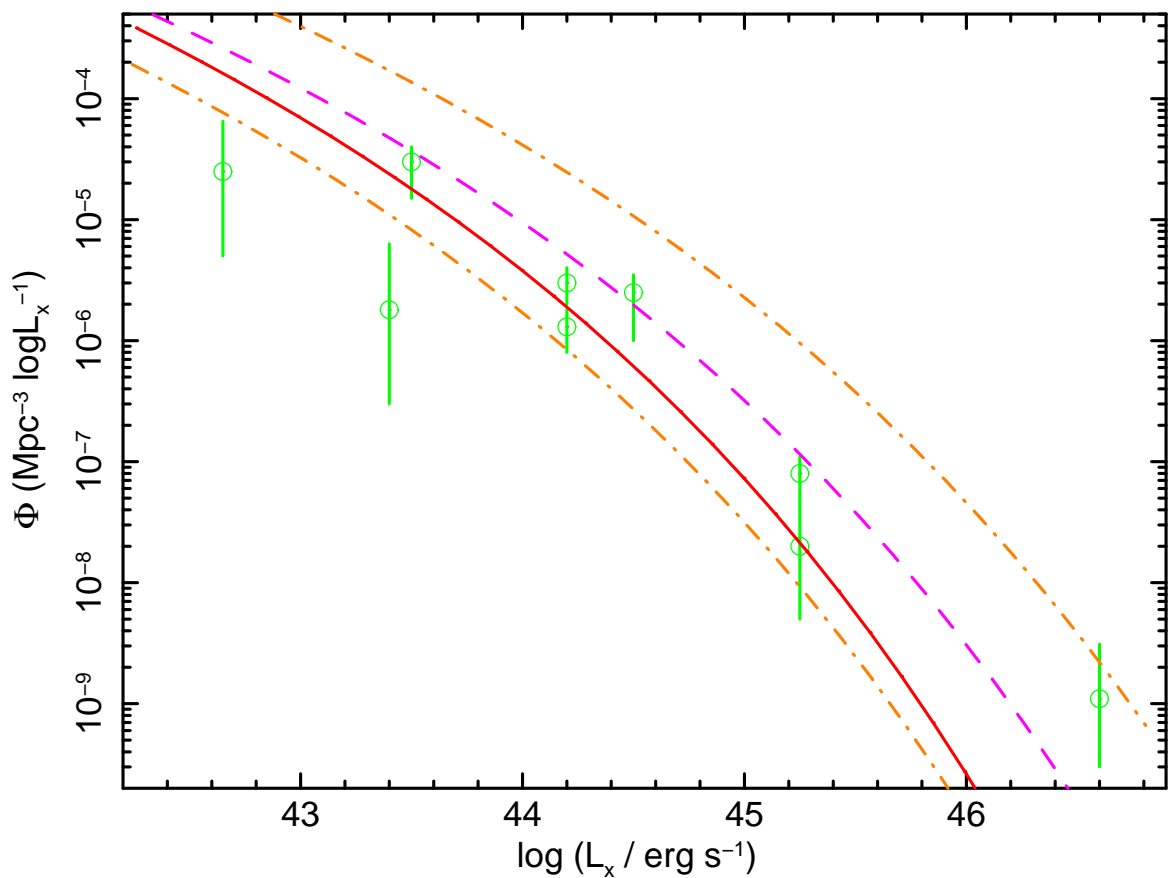}
\caption{AGN LFs in the $2-10$ keV band at $z=4.5$. The data
points with error bars are the observational results at
$4.0<z<5.0$ given by \citet{ueda14}. The denotations of solid, dashed, and
two dotted-dashed lines are the same as those in Figure 4.
\label{fig6}}
\end{figure}

\begin{figure}
\includegraphics[angle=270,scale=0.65]{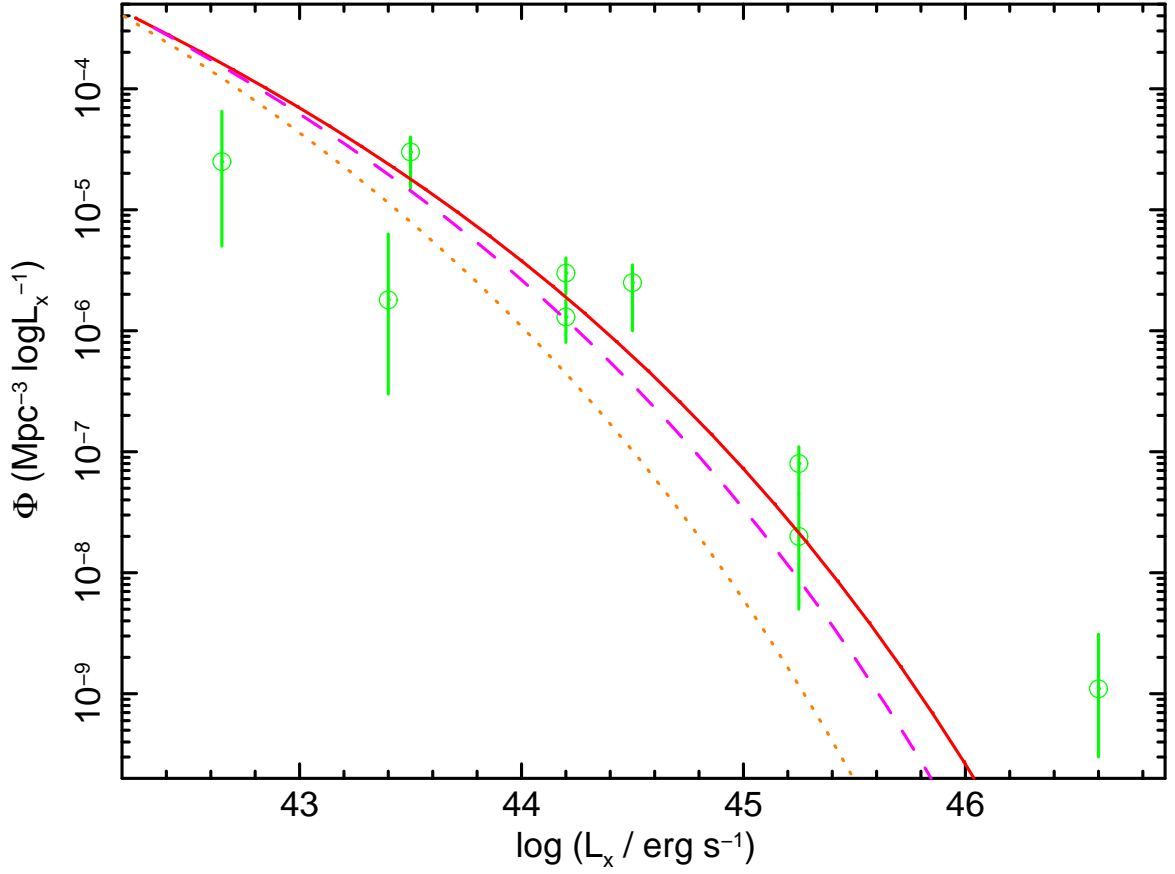}
\caption{AGN LFs in the $2-10$ keV band at $z=4.5$ (solid line), $z=5$
(dashed line), and $z=6$ (dotted line), respectively. The $M_{\rm{BH}}-\sigma$
relation given by \citet{robertson06} is considered to reproduce
these AGN LFs. Data points
with error bars are the observational AGN LFs given by
\citet{ueda14} at $4.0<z<5.0$.
\label{fig7}}
\end{figure}

\begin{figure}
\includegraphics[angle=270,scale=0.65]{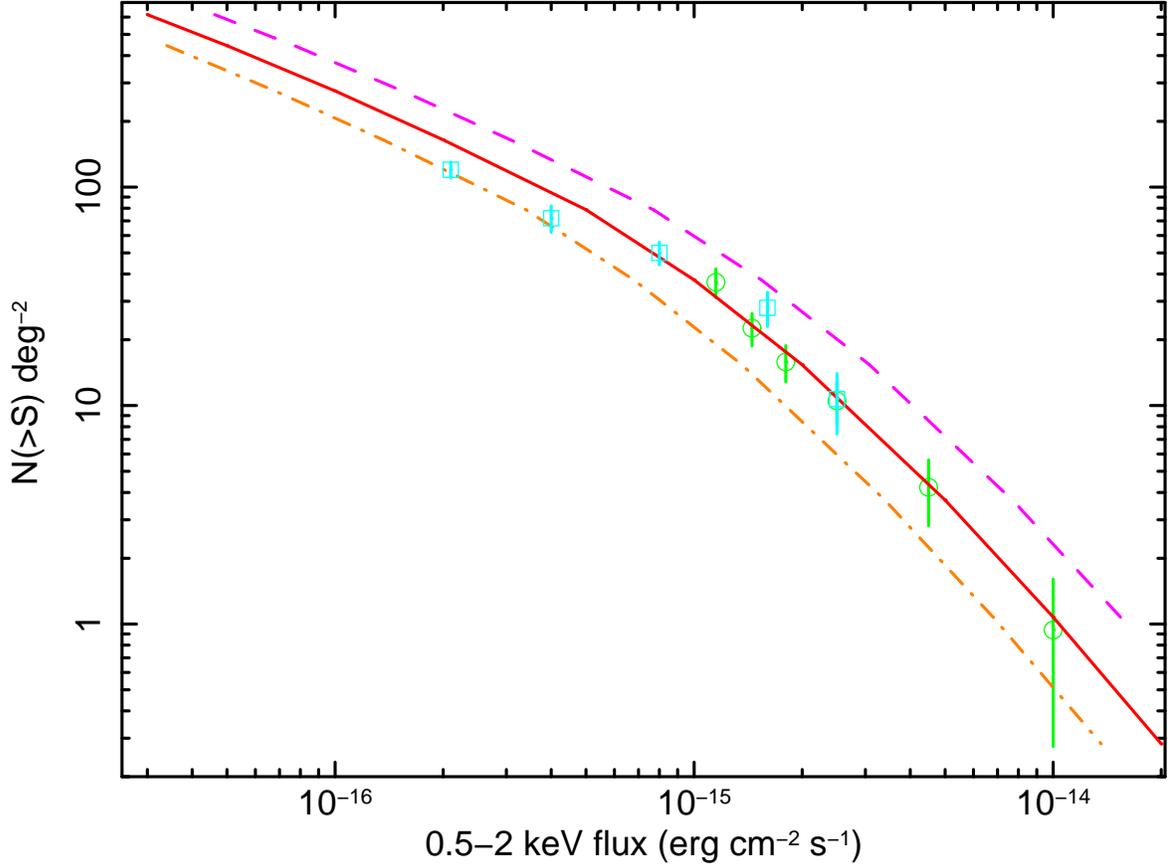}
\caption{AGN number counts in the $0.5-2$ keV band at $z>3$.
Data points with error bars denoted by circles and squares are those
observational results from the XMM-COSMOS survey \citep{brusa09} and the {\it{Chandra}}-COSMOS survey \citep{civano11}, respectively.
Solid, dashed, and dotted-dashed lines denote our results
with $\rm{N_H}$ values of
$5.0\times 10^{22}~\rm{cm^{-2}},~1.0\times 10^{22}~\rm{cm^{-2}},~and~
1.0\times 10^{23}~\rm{cm^{-2}}$, respectively.
The $M_{\rm{BH}}-\sigma$ relation from the numerical simulations given
by \citet{robertson06} is considered in our calculations. \label{fig8}}
\end{figure}

\begin{figure}
\includegraphics[angle=270,scale=0.65]{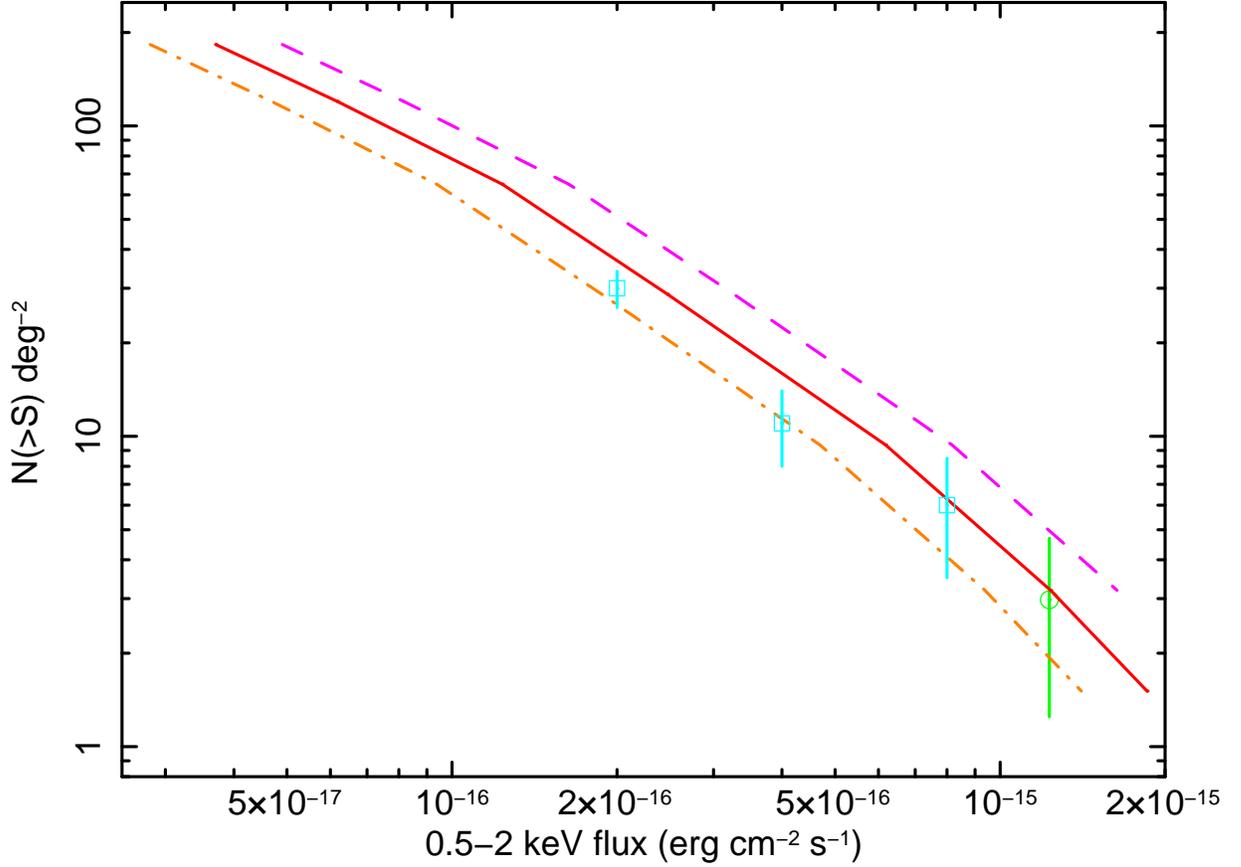}
\caption{AGN number counts in the $0.5-2$ keV band at $z>4$.
Data points with error bars denoted by circles and squares are
the observational results from the XMM-COSMOS survey \citep{brusa09} and
the {\it{Chandra}}-COSMOS survey \citep{civano11}, respectively. The
denotations of solid, dashed, and dot-dashed lines are the same
as those in Figure 8. The $M_{\rm{BH}}-\sigma$ relation from the numerical simulations given
by \citet{robertson06} is considered in our calculations.\label{fig9}}
\end{figure}

\begin{figure}
\includegraphics[angle=270,scale=0.65]{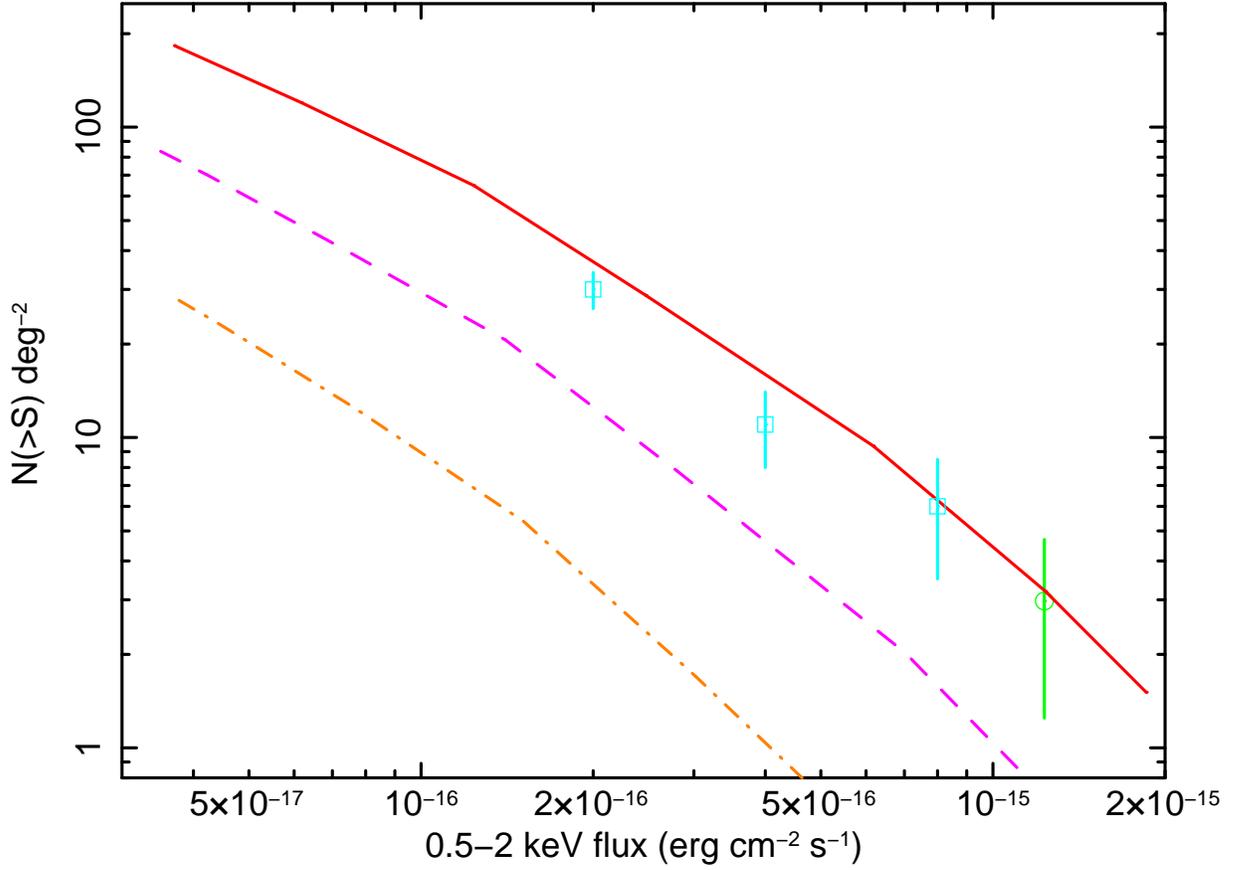}
\caption{AGN number counts in the $0.5-2$ keV band at $z>4$ (solid line),
$z>5$ (dashed line), and $z>6$ (dotted-dashed line), respectively. The theoretical $M_{\rm{BH}}-\sigma$
relation given by \citet{robertson06} and the absorption of
$\rm{N_H}=5.0\times 10^{22}~\rm{cm^{-2}}$ are considered.
Data points with error bars denoted by circles and squares are the same as
those in Figure 9.
\label{fig10}}
\end{figure}

\begin{figure}
\includegraphics[angle=270,scale=0.65]{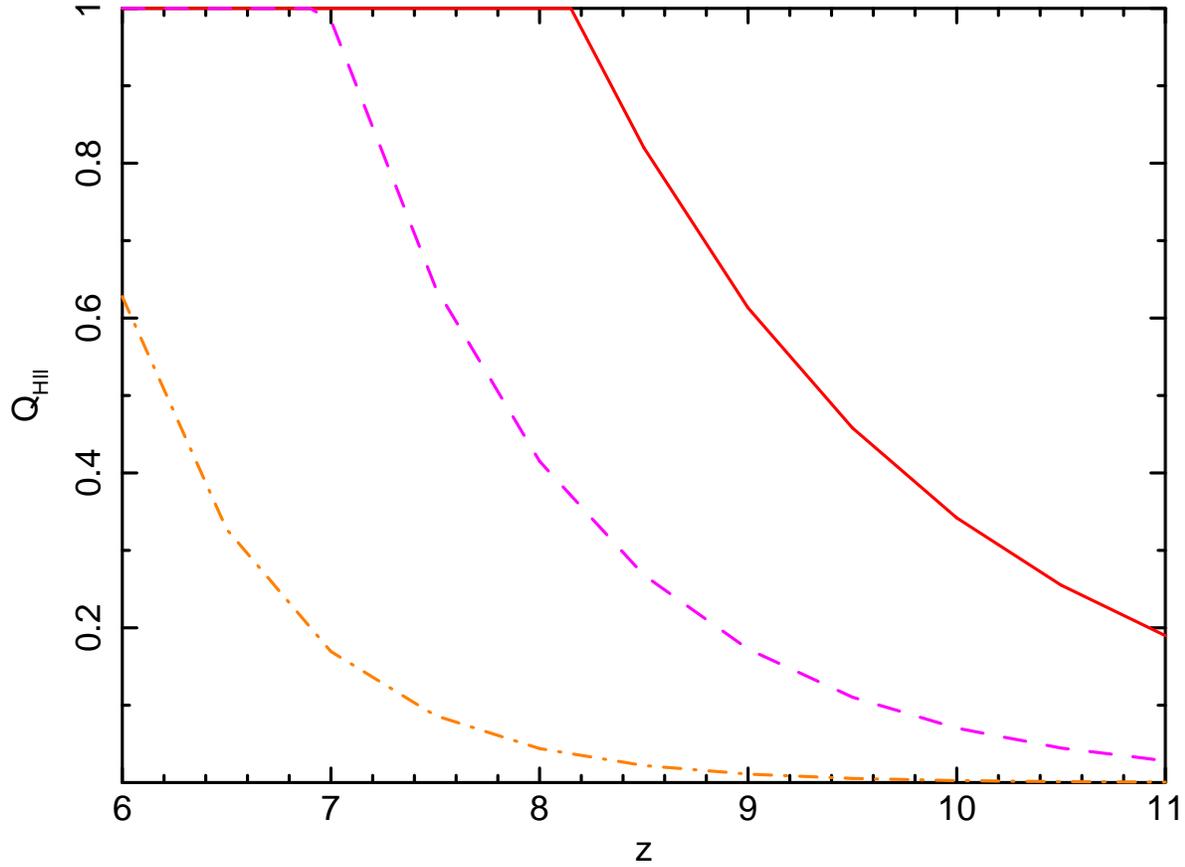}
\caption{Evolution of filling factor $Q_{\rm{HII}}$. Solid, dashed
and dotted-dashed lines denote our results from the contributions
of BHs with mass ranges of $10^6-10^7$ $M_\odot$, $10^7-10^8$
$M_\odot$, and $10^8-10^9$ $M_\odot$, respectively. \label{fig11}}
\end{figure}

\begin{figure*}
\vspace{5pt}
\includegraphics[angle=270,scale=0.28]{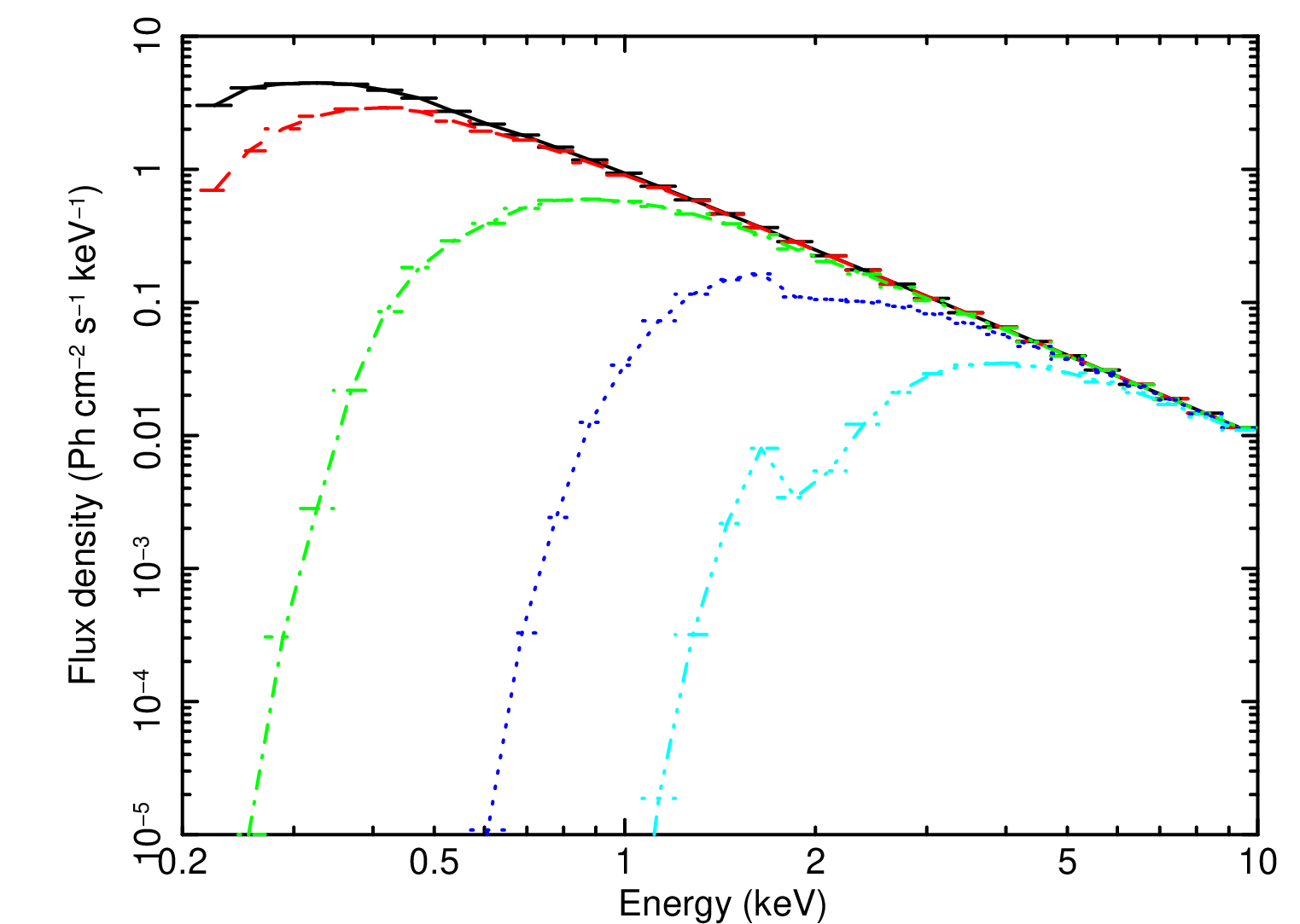}
\includegraphics[angle=270,scale=0.28]{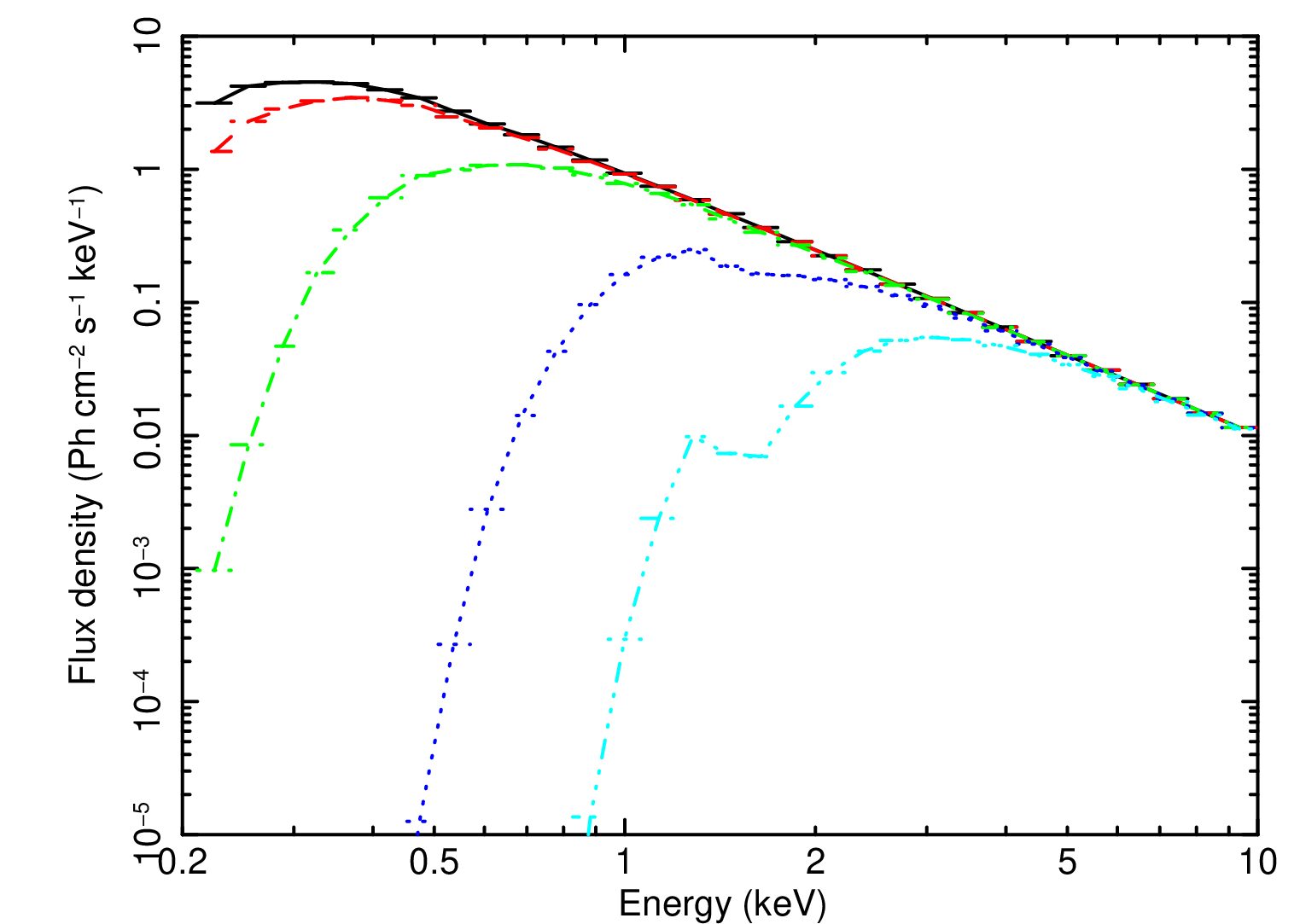}\\
\includegraphics[angle=270,scale=0.28]{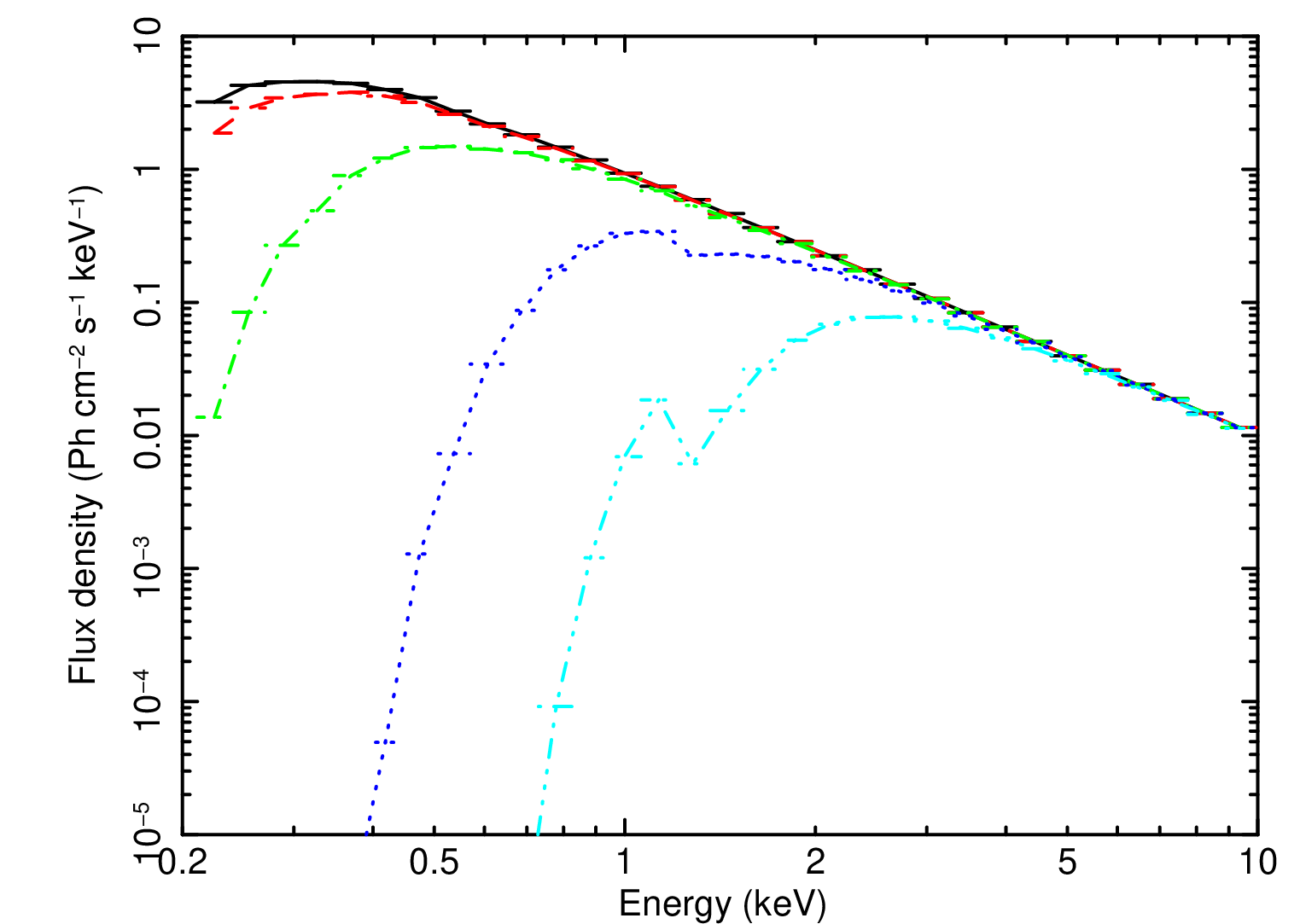}
\includegraphics[angle=270,scale=0.28]{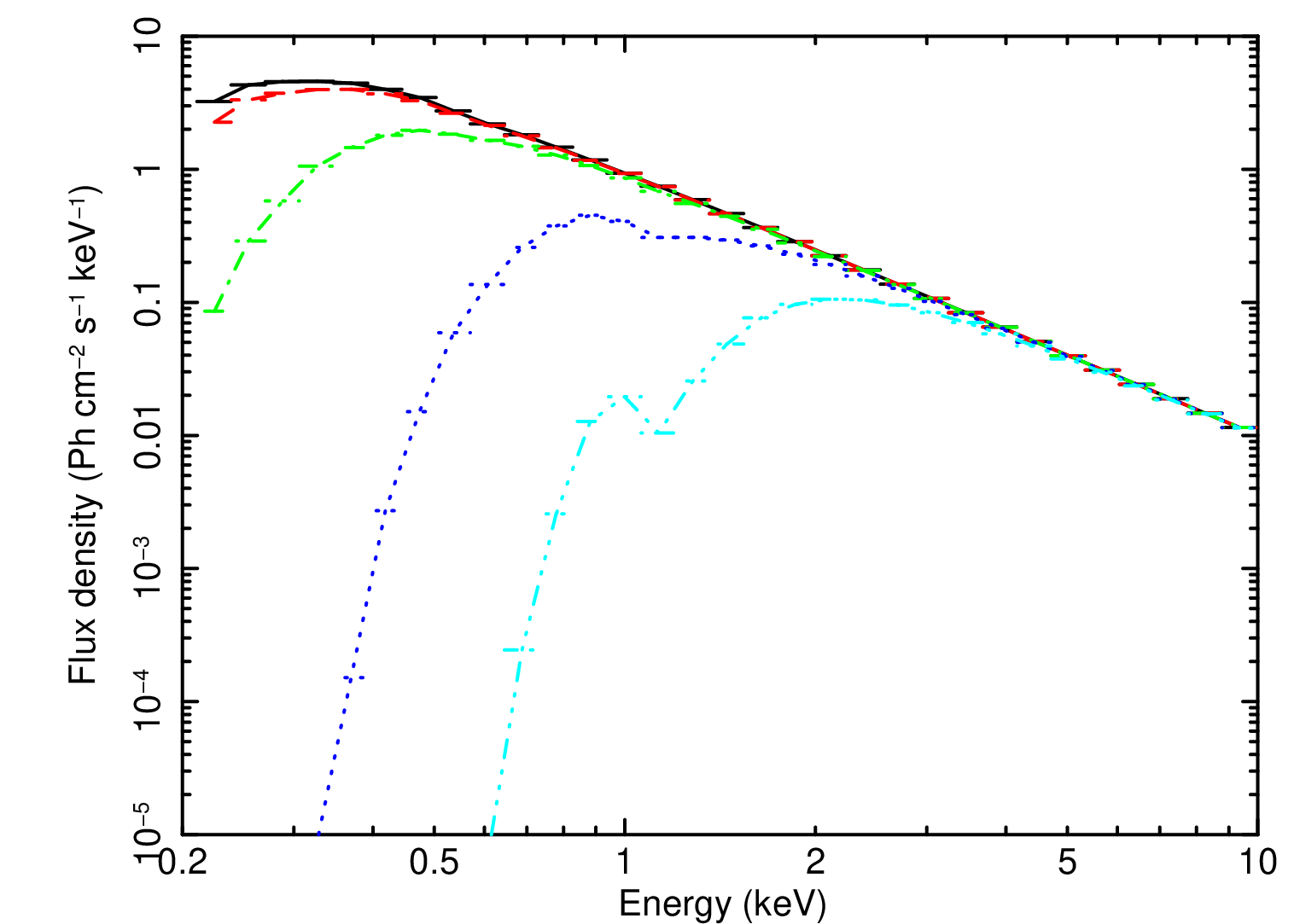}
\caption{Absorbed spectral templates in the X-ray band. Several cases of different $\rm{N_H}$ values
($2.9\times 10^{20}~\rm{cm^{-2}},~5.0\times 10^{21}~\rm{cm^{-2}},~5.0\times 10^{22}~\rm{cm^{-2}},~5.0\times 10^{23}~\rm{cm^{-2}},
~\rm{and}~2.5\times 10^{24}~\rm{cm^{-2}}$) are denoted by a solid line, dashed line, dotted-dashed line,
dotted line, and dotted-dotted-dotted-dashed line, respectively. Top-left panel: spectra at $z=3$; Top-right panel:
spectra at $z=4$; Bottom-left panel: spectra at $z=5$; Bottom-right panel: spectra at $z=6$.}
\label{appenfig}
\end{figure*}

\begin{figure*}
\vspace{5pt}
\includegraphics[angle=270,scale=0.28]{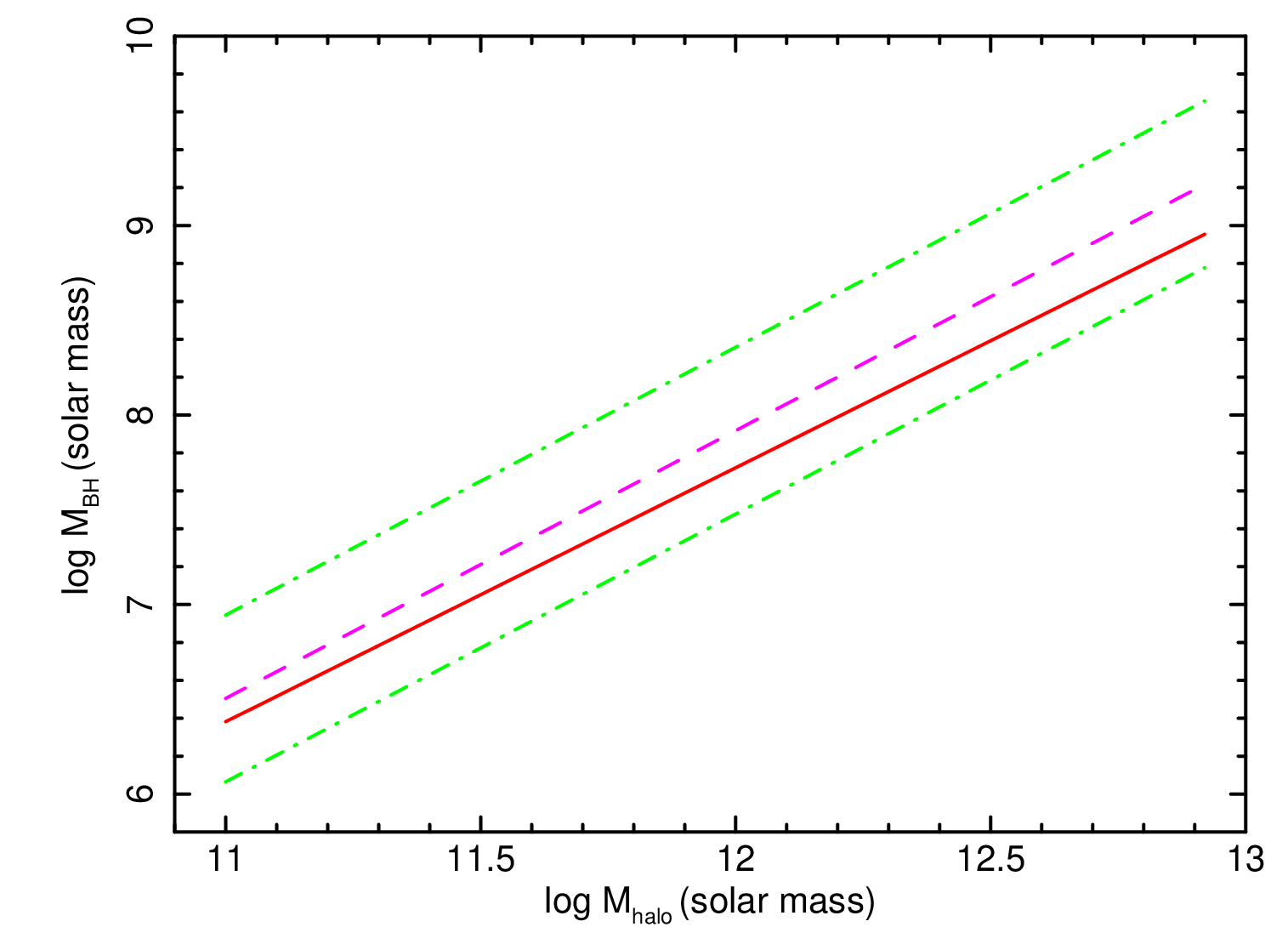}
\includegraphics[angle=270,scale=0.28]{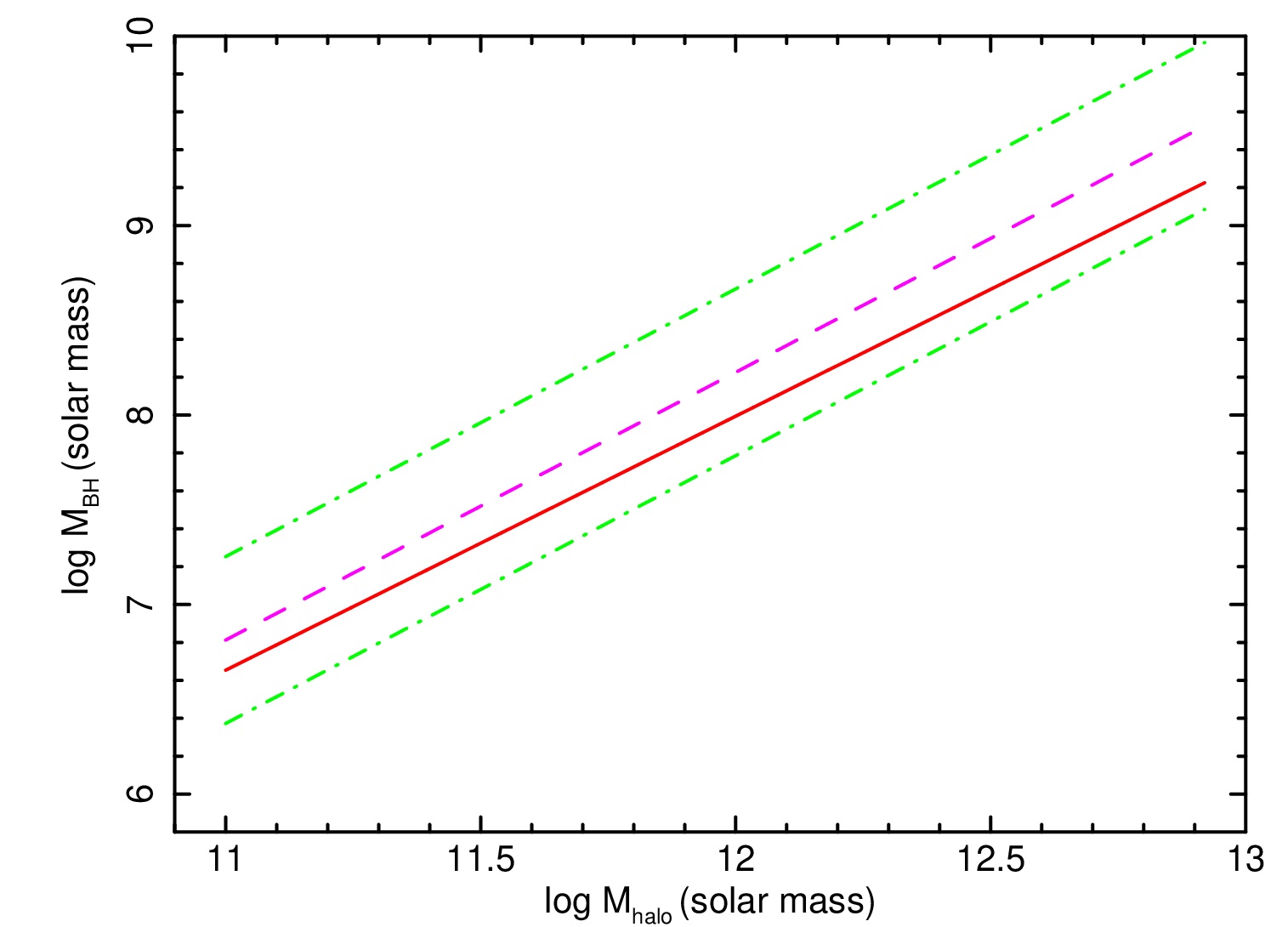}\\
\includegraphics[angle=270,scale=0.28]{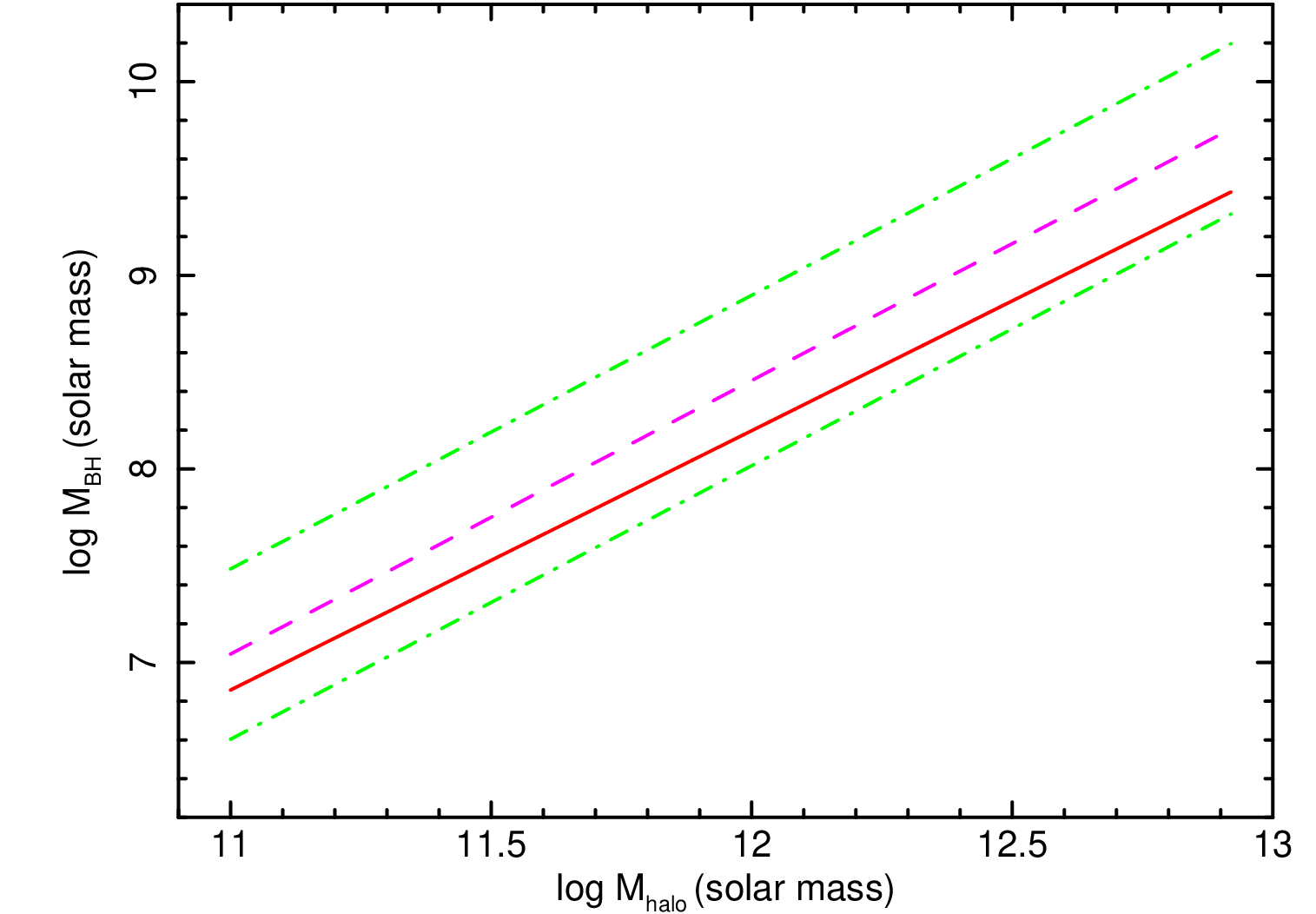}
\includegraphics[angle=270,scale=0.28]{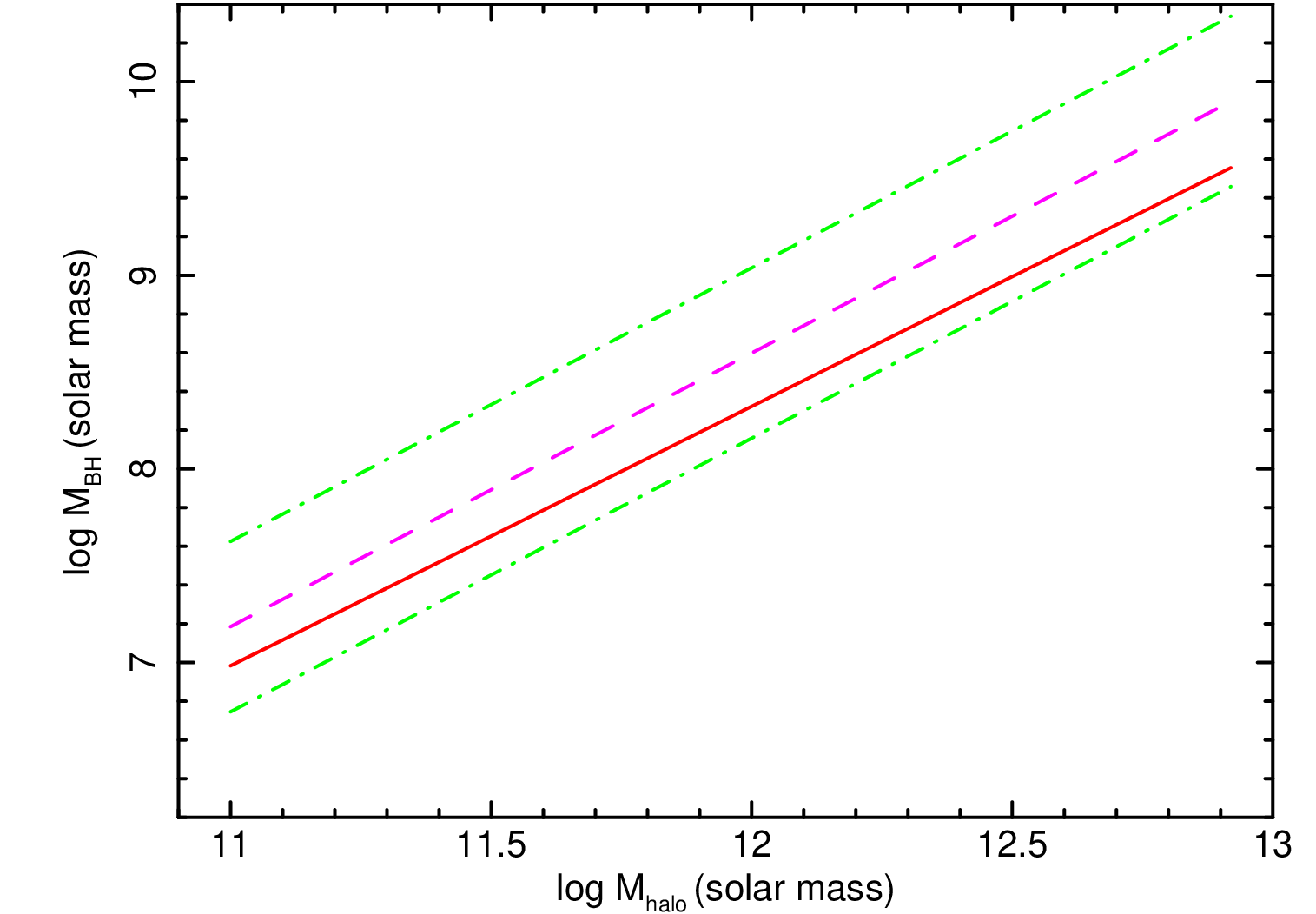}
\caption{BH$-$Halo mass relation.
The solid line denotes our result, adopting the $M_{\rm{BH}}-\sigma$ relation from the
numerical simulation given by \citet{robertson06}. The dashed line
denotes our result, adopting the observational $M_{\rm{BH}}-\sigma$ relation given by \citet{gultekin09}. The two dotted-dashed lines denote the lower and upper limits of our results in the case in which the intrinsic scatter of the observational $M_{\rm{BH}}-\sigma$ relation given by
\citet{gultekin09} is considered.
top-left panel: relation at $z=4$; top-right panel:
relation at $z=6$; bottom-left panel: relation at $z=8$; bottom-right panel: relation at $z=9.5$.}
\label{appenfig}
\end{figure*}

\begin{figure*}
\vspace{5pt}
\includegraphics[angle=270,scale=0.28]{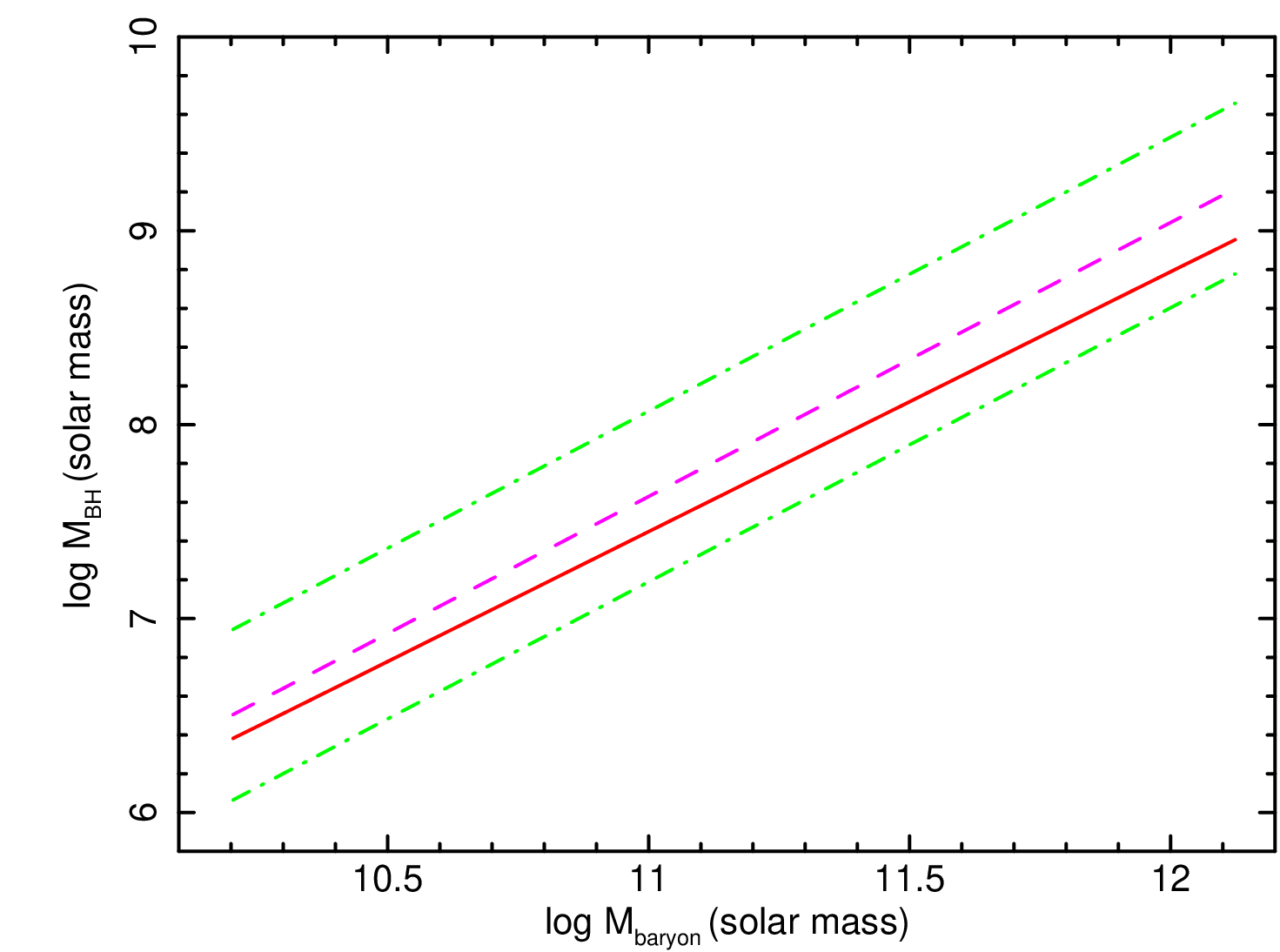}
\includegraphics[angle=270,scale=0.28]{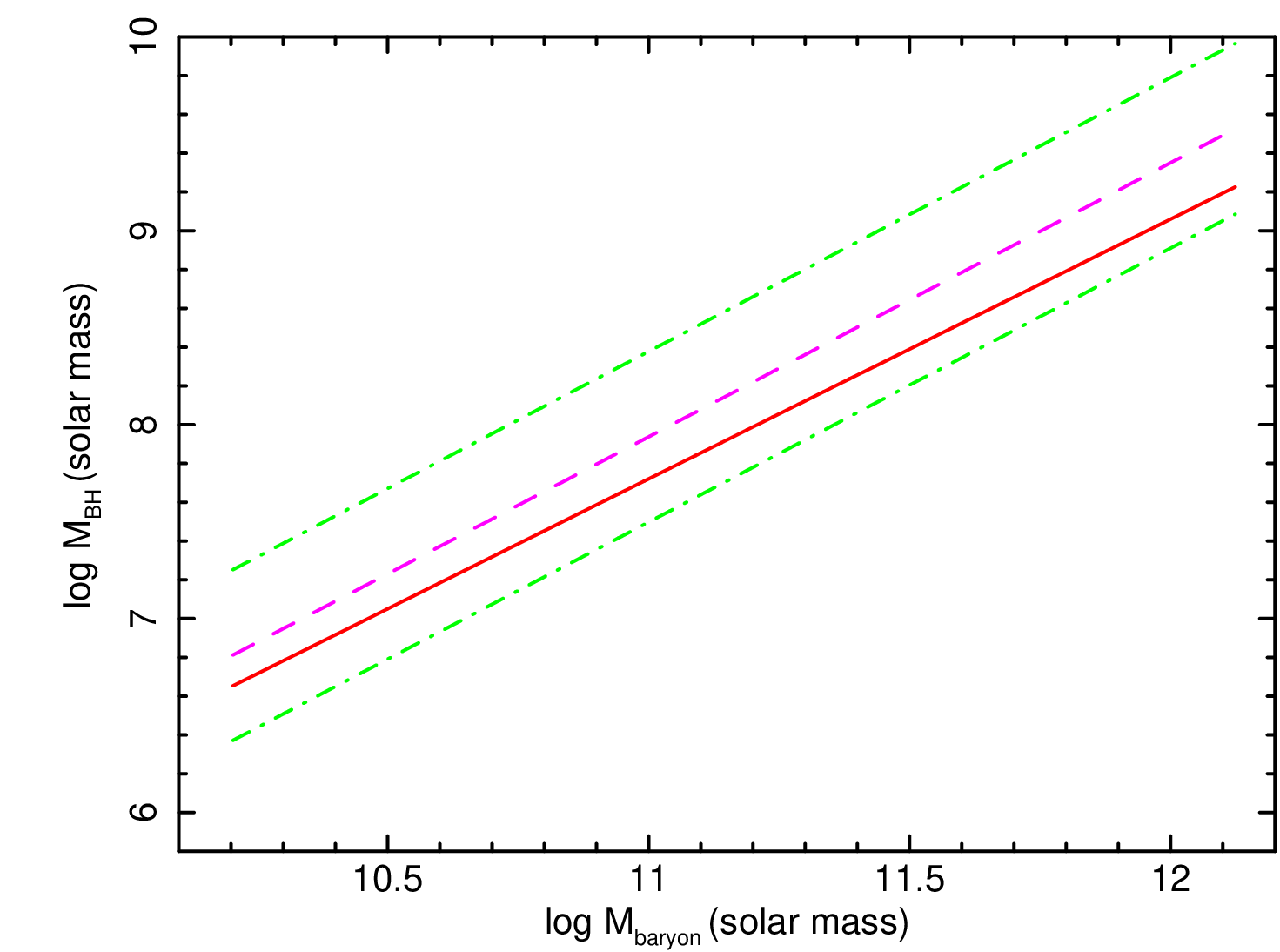}\\
\includegraphics[angle=270,scale=0.28]{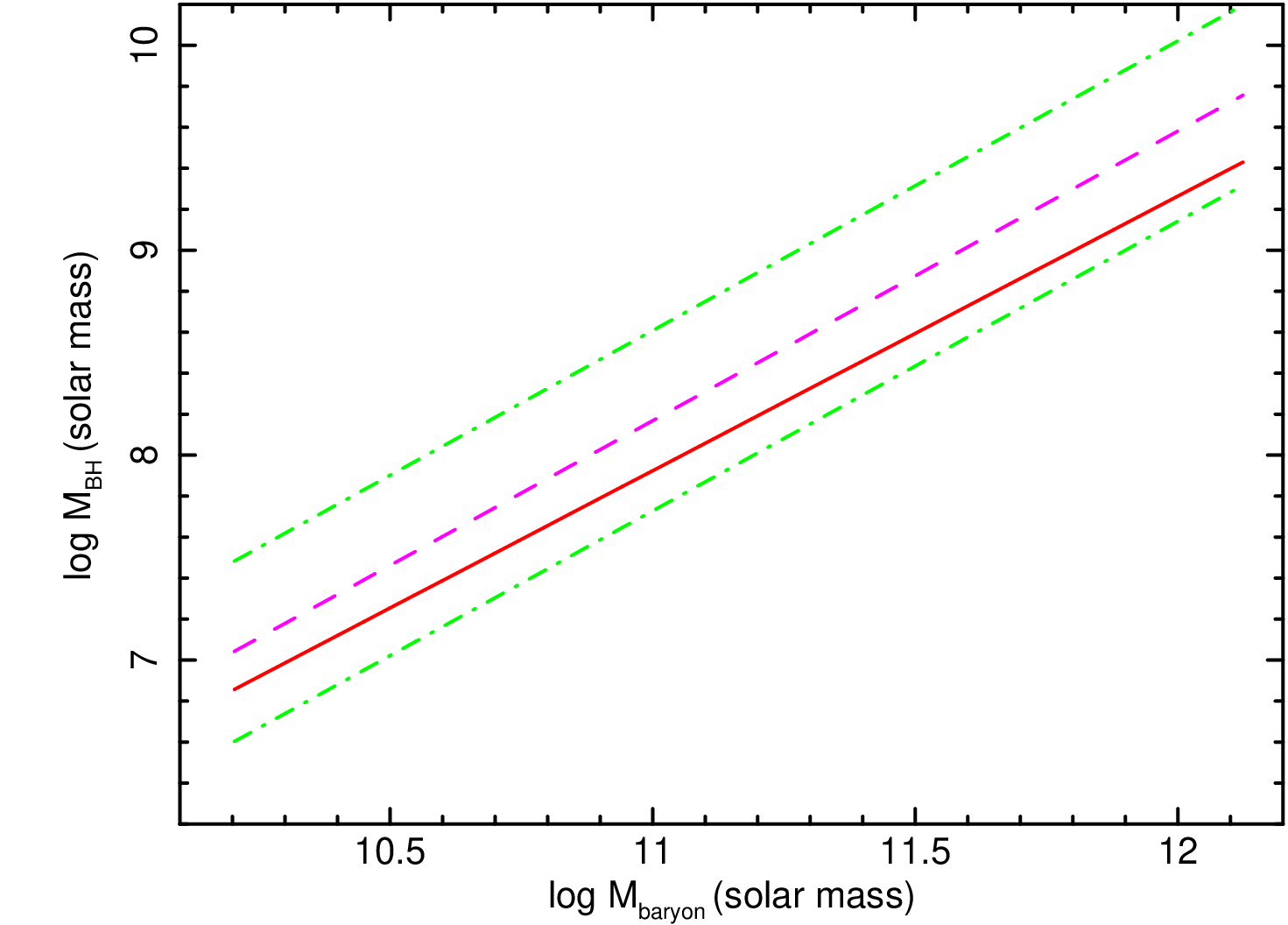}
\includegraphics[angle=270,scale=0.28]{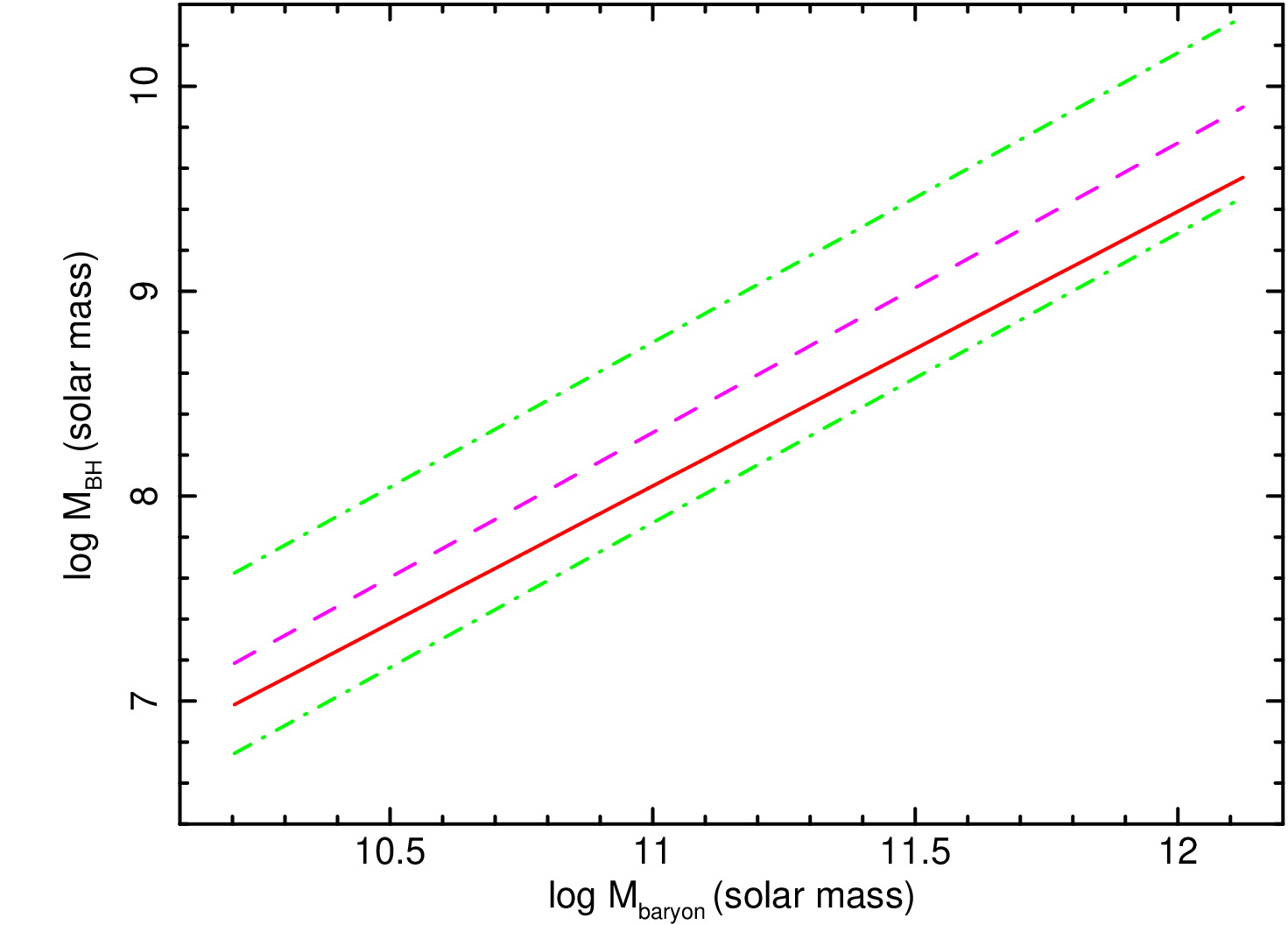}
\caption{BH$-$baryon mass relation.
The denotations of solid,
dashed, and two dotted-dashed lines are the same as those in Figure 13.
top-left panel: relation at $z=4$; top-right panel:
relation at $z=6$; bottom-left panel: relation at $z=8$; bottom-right panel: relation at $z=9.5$.}
\label{appenfig}
\end{figure*}

\begin{table*}
\begin{center}
\caption{Parameters adopted in the Calculation Processes of the AGN LFs and AGN Number
Counts \label{tab:1}}
\end{center}
\begin{tabular}{cccccc}
\hline \hline
\footnotesize{Parameter} & \footnotesize{Optical Band} & \footnotesize{X-ray Band} \\
\hline
\footnotesize{Eddington ratio} & $\lambda=0.25(M_{\rm{BH}}/2\times 10^9)^{0.5}$ & $\lambda=0.25(M_{\rm{BH}}/2\times 10^9)^{0.5}$ \\
\footnotesize{AGN lifetime (years)} & \footnotesize{$2\times 10^7$} & \footnotesize{$9\times 10^7$}   \\
\footnotesize{$\alpha$ in $M_{\rm{BH}}-\sigma$ relation \citep{robertson06}}& \footnotesize{$8.01-0.138log(1+z)$}  & \footnotesize{$8.01-0.138log(1+z)$} \\
\footnotesize{$\beta$ in $M_{\rm{BH}}-\sigma$ relation \citep{robertson06}}& \footnotesize{4.02}  & \footnotesize{4.02}   \\
\footnotesize{$\alpha$ in $M_{\rm{BH}}-\sigma$ relation \citep{gultekin09}}& \footnotesize{8.12}  & \footnotesize{8.12}   \\
\footnotesize{$\beta$ in $M_{\rm{BH}}-\sigma$ relation \citep{gultekin09}}& \footnotesize{4.24}   & \footnotesize{4.24}   \\
\hline \hline
\tablecomments{
The exception case is for the optical AGN LF at $z=6$ given by \citet{willott10b}: $\lambda=2.0(M_{\rm{BH}}/2\times 10^9)^{0.5}$ and AGN lifetime of $2\times 10^5$ yrs.
$\alpha$ and $\beta$ are defined as $\rm{log}(M_{\rm{BH}}/M_\odot)=\alpha+\beta\rm{log}(\sigma/200~\rm{km~s^{-1}})$.}
\end{tabular}
\end{table*}

\end{document}